\documentstyle[epsf]{mn}  
\newcommand{\NOTTING}{School  of Physics and Astronomy, 
University of Nottingham, University Park, Nottingham, NG7 2RD, UK}
\newcommand{\USNO}{Dept. of the Navy, USNO,
3450 Massachusetts Ave NW, Washington, DC 20392-5420, USA}
\newcommand{\USRA}{Universities Space Research Association,
300 D Street, SW, suite 801, Washington, DC 20024-4703, USA}
\newcommand{\RSUN}{\mbox{$R_0 \;$}}
\newcommand{\RSUNn}{\mbox{$R_0$}}
\newcommand{\VSUN}{\mbox{$\Theta_0 \;$}}
\newcommand{\VSUNn}{\mbox{$\Theta_0$}}

\newcommand{\Sgas  }{\mbox{$\Sigma_g$}}
\newcommand{\Sstr}{\mbox{$\Sigma_*$}}
\newcommand{\SKG}{\mbox{$\Sigma^{1.1}_{tot}$}}
\newcommand{\SHkg}{\mbox{$\Sigma^{1.1}_{h}$}}
\newcommand{\sgas  }{\mbox{$\sigma_g$}}
\newcommand{\sgasp }{\mbox{$\sigma'_g$}}
\newcommand{\FWHM  }[1]{\mbox{${\rm FWHM{#1}}$}}
\newcommand{\FWHMrp}[2]{\mbox{${\rm FWHM^{#2}{#1}}$}}

\newcommand{\kms}{$\mbox{km\,s}^{-1}$}
\newcommand{\kmskpc}{$\mbox{km\,s}^{-1}\,\mbox{kpc}^{-1}$}
\newcommand{\hd}{\mbox{$h_{\rm d}$}}
\newcommand{\ze}{\mbox{$z_{\rm e}$}}
\newcommand{\Kz}[1]{\mbox{$K_{\rm z}{#1}$}}
\newcommand{\qHI}{\mbox{$q_{{\small \rm \HI}}$}}
\newcommand{\qdT}{\mbox{$q_{{\small \rm 1.1}}$}}

\newcommand{\Rc}{\mbox{$R_{\rm c}$}}
\newcommand{\rhoN}{\mbox{$\rho_0$}}
\newcommand{\rhoROT}[1]{\mbox{$\frac{-1}{2\pi\rm{G}}\frac{V{\rm #1}}{R}\frac{dV{\rm #1}}{dR}$}}

\newcommand{\Ht}{\mbox{${\rm H_2}$}}
\newcommand{\HI}{\mbox{{\rm H \footnotesize{I} }}}
\newcommand{\HII}{\mbox{{\rm H \footnotesize{II} }}}

\newcommand{\LsunB}[1]{\mbox{$L_{\odot,#1}$}}
\newcommand{\LSpcsq}{\mbox{$L_{\odot}\,{\rm pc}^{-2}$}}
\newcommand{\MSpcsq}{\mbox{$M_{\odot}\,{\rm pc}^{-2}$}}
\newcommand{\MSpccub}{\mbox{$M_{\odot}\,{\rm pc}^{-3}$}}

\newcommand{\Ups}[1]{\mbox{$\Upsilon_{{\rm #1}}$}}

\newcommand{\rtp}[1]{\mbox{$^{#1}$}}

\newcommand{\pmt}{\mbox{$\pm$}}

\def\lesssim{\mathrel{\hbox{\rlap{\hbox{\lower4pt\hbox{$\sim$}}}\hbox{$<$}}}}
\def\gtrsim{\mathrel{\hbox{\rlap{\hbox{\lower4pt\hbox{$\sim$}}}\hbox{$>$}}}}
\let\la=\lesssim
\let\ga=\gtrsim

\newcommand{\nid}{\noindent}
\newcommand{\mVskip}[1] {\vspace*{ #1}}
\newcommand{\negSMLskip}{\vspace*{ -3mm}}

\newcommand{\etal}{\mbox{{\it et al. }}}

\def\apj{ApJ}
\def\apjl{ApJ}

\def\aj{AJ}
\def\mnras{MNRAS}
\def\aap{A\&A}

\newcommand{\ptn}{\phantom{0.0000}}

\begin{document}

\title[Luminous and Dark Matter in the Milky Way]
      {Luminous and Dark Matter in the Milky Way}

\author[R.P. Olling, M.R. Merrifield]
  {Rob P. Olling$^{A,B}$\thanks{E-mail: olling@usno.navy.mil},
  Michael R. Merrifield$^{C}$\thanks{E-mail: Michael.Merrifield@Nottingham.ac.uk}  \\
$^A$\USNO\\
$^B$\USRA\\
$^C$\NOTTING }

\date{Accepted for Publication}

\maketitle

\begin{abstract} 

Axisymmetric models of the Milky Way exhibit strong interrelations
between the Galactic constants [the Sun's distance to the Galactic
centre (\RSUNn), and the local rotation speed (\VSUNn)], the local
stellar columndensity ($\Sigma_*(R_0)$) and the shortest-to-longest
axis ratio of the dark matter halo ($q$).  In this paper we present
simple analytical approximations that allow for an efficient search
through the vastness of parameter space, and apply this approximation
to investigate the consequences of the uncertain gaseous velocity
dispersion (\sgas) on the constraints imposed by the thickness of the
Milky Way's gas layer.  The extra degree of freedom does not
significantly alter the conclusions drawn in a previous paper on the
shape of the Milky Way's dark matter halo.  A significant contribution
to the total gas pressure by cosmic rays and magnetic fields beyond
the optical disk is thus ruled out.  We find that the Milky Way's dark
halo is close to spherical if $\RSUN \ga 7.1$ kpc, while a
significantly flattened dark matter halo is only possible if our
distance to the Galactic centre is smaller than $\sim$ 6.8 kpc.

Thus, if \RSUN is larger than $\sim$7 kpc, or \VSUN $\ga$ 170 \kms, we
can rule out two dark matter candidates that require a highly flattened
dark matter halo: 1) decaying massive neutrinos; and 2) a disk of cold
molecular hydrogen.

It is only possible to construct a self-consistent axisymmetric
model of the Galaxy based on the IAU-recommended values for the
Galactic constants (\RSUN = 8.5 kpc, \VSUN = 220 \kms) in the unlikely
case that the effective gaseous velocity dispersion is $\sim$19\%
larger than observed, {\em and} if the local stellar columndensity is
less than about 18 \MSpcsq.  If we assume that the halo is oblate and
a value of \Sstr\ of 35 \pmt 5 \MSpcsq \cite{kKgG89b}, we can rule out
Galactic models with $\RSUN \ga 8.0$ kpc and $\VSUN \ga 200$ \kms.

Combining the best kinematical and star-count estimates of \Sstr, we
conclude that \Sstr\ probably lies between 25 and 45 \MSpcsq.  We find
that Kuijken \& Gilmore's (1991) determination of the columndensity of
matter within 1.1 kpc of the plane is robust and valid over a wide
range of Galactic constants.

Our mass models show that, largely due to the uncertainty in the
Galactic light distribution, the dark matter density in the Galactic
centre is uncertain by up to three orders of magnitude.  In the Solar
neighbourhood this uncertainty is much reduced: our models imply a dark
matter density of some 0.42 GeV/c\rtp{2} per cubic centimetre, or (11
\pmt 5) m\MSpccub -- roughly 15\% of the total mass density. 

\end{abstract}

\begin{keywords}
Galaxy: structure          - Galaxy: kinematics and dynamics - 
Galaxy: solar neighborhood - Galaxy: fundamental parameters -
Galaxy: stellar content
\end{keywords}

\mVskip{-5mm}
\section{Introduction}
\label{sec:Introduction}

The observational fact that rotation curves of external galaxies are
rather flat in their outer parts
\cite{RFT80,aB81,smK87,kgB87,sCjvG91,ahB92,PSS96} indicates the presence
of unseen matter in those galaxies\footnote{See McGaugh \& de Blok
(1998) for a review of the alternative hypothesis that the law of
gravity has to be modified instead.}.  A problem which is specific for
the Milky Way is that we do not know the shape of the Galactic rotation
curve (RC).  The slope of the RC depends on the assumed values of the
Galactic constants \cite{rOmM98b98c}.  But dark matter is required
whatever the values of the Galactic constants.  A fit to the observed
rotation curve can be used to yield the stellar mass-to-light ratio
\Ups{d}, and the dark halo parameters, albeit with large uncertainties. 
Rather than using this fitting procedure, we adopt an analytical
approximation in which there is only one free parameter: the degree to
which the disk is maximal ($\gamma$), from which \Ups{d}\ and the dark
halo parameters follow (see Olling 1995 for details).  The parameter
$\gamma$ is preferred for dynamical modeling purposes as it is bound
between 0 and 1 for a zero mass and a full-fledged disk, respectively. 
For example, in the popular ``maximum-disk'' hypothesis \cite{vAS86},
the amplitude of the stellar rotation curve equals (85 \pmt 10)\% of the
observed rotation speed \cite{pdS97}, so that $\gamma=0.85$.  In
contrast, Bottema (1993) used stellar velocity dispersion measurements
to ``weigh'' stellar disks, and concluded that they are sub-maximal,
with $\gamma = 0.63 \pm 0.1$.  The situation is similarly indeterminate
for the Milky Way: the Kuijken \& Gilmore (1989, hereafter referred to
as KG89) model implies $\gamma \sim 0.5$, while more recent models with
lower rotation speed and shorter scale-lengths can be close to maximal
($\gamma = 0.85 \pm 0.1$; Sackett 1997).  As a result of the uncertain
stellar mass-to-light ratio, the dark halo parameters are very ill
determined for most galaxies \cite{egvAS86,LF89,rpO95}.  The Milky Way
is no different: the combined uncertainty in the local disk mass and the
stellar scale-length introduces an uncertainty of over three orders of
magnitude in the central dark matter (DM) density of the Milky Way, and
about an order of magnitude uncertainty in its core radius (\S
\ref{sec:Mass_Models};Dehnen \& Binney 1998).  In the Solar neighborhood
the situation is less dramatic, although no consensus exists on the
local volume density of dark matter [$\rho_{DM}$ or the Oort limit; see
\S \ref{sec:Mass_Models} and Cr\'{e}z\'{e} \etal (1998) for a recent
review].  However, the local dark matter density is important for many
astrophysical problems.  For example, if the dark matter comprises
elementary particles like neutralinos, axions, neutrinos, gravitinos
etc., their expected detection rate is proportional to the DM density. 
Likewise, if the dark halo is made up of massive compact halo objects
(MACHOs), the event rate for gravitational lensing depends on the
integrated dark matter density along the line of sight towards the lens. 
Thus, observational signatures of the Milky Way's dark matter
distribution like micro-lensing time scales and optical depths
\cite{GGT95}, and expected neutralino annihilation rate \cite{lBpUjB98}
depend on the Galactic dark matter density distribution and hence the
assumed values for the Galactic constants. 

   In a previous paper \cite{PaperIhence} we determined the dark
matter density in the Solar neighbourhood at
$R$\footnote{Throughout this paper we use cylindrical coordinates
with $R$ the Galactocentric distance, and $z$ the distance of the
plane.}=\RSUN  and around $R\sim2R_0$ to infer the minor to major axial
ratio $(q=c/a)$, or shape, of the dark matter halo of the Milky
Way. In the present paper we investigate the reliability of several of
the assumptions made in Paper~I, and find that relaxing these
assumptions does not greatly change the conclusion of Paper~I: the
shape of the dark matter halo of the Milky Way is probably rather
round. Before going into more detail, let us review some of the
difficulties which arise when one tries to determine $\rho_{DM}$.

   First, large values of the local Galactic rotation speed (\VSUNn)
result in large DM densities, while low rotation speeds require small
DM densities.  Second, since the {\em shape} of the Galactic rotation
curve depends on the value of our distance to the Galactic center
\cite{rOmM98c}, the amount of dark matter increases with \RSUN at
constant \VSUNn.  And finally, more highly flattened DM halos have
larger midplane densities \cite{rpO95}.

We use two sets of observations to constrain the midplane dark matter
density.  First, stellar kinematical data provides a measure of the
total columndensity within 1.1 kpc of the plane (\SKG; cf. Kuijken \&
Gilmore 1991).  The dark matter density follows after subtracting the
luminous components and dividing by the scale-height.  This method
yield ambiguous results because uncertainties in the surface density
of stellar matter (\Sstr) translate into a similar uncertainty of the
DM density.  Thus, low values of \Sstr\ require more dark matter and
hence a more highly flattened dark halos at constant \RSUN and \VSUNn.

Second, the rate at which the thickness of the gas layer increases
with radius (``flaring'') is a measure of $\rho_{DM}$.  Assuming, for
now, a hydrostatic balance between internal pressure and gravity, it
follows that an increasing gas layer width is evidence for a
decreasing midplane density.  Large dark matter densities result in
thin gas layers, while low densities yield a thicker gas disk.  A
larger DM density, due to either a larger \VSUN and/or \RSUN or a
smaller $q$, results in a thinner gas layer.  However, such a thinner
gas layer would also occur if the actual gas pressure --or
equivalently, the gaseous velocity dispersion, \sgas-- is smaller than
assumed. Hence the significant correlations between the assumed values
of the Galactic constants, \Sstr\ and \sgas\ and the inferred shape of
the Milky Way's dark matter halo referred to above.

  In practice, the stellar kinematical data impose correlations
between \VSUN and $q$ at the Solar circle, while the observed \HI\
flaring does so at $R\sim (2\pm0.25)\RSUN$. At these large radii the
stellar disk has vanished so that the potential is dominated by the
gaseous self-gravity and the dark matter.

   Unfortunately, the Galactic constants are ill-determined
\cite{KLB86,mjR93,rOmM98aPI}, and this has consequences for analyses
that depend on \RSUN and \VSUNn.  In an ideal world, parameters such as
the Galactic constants, the rotation curve [$\Theta(R)$], the
scale-length of the optical disk (\hd), the local stellar column
density, the gaseous velocity dispersion, and so forth, are not only
measured, but also have normal errors.  In that case one could determine
quantities that depend on these parameters, such as the DM density or
the halo's shape, by comparing the model and observed parameters in a
$\chi^2$ sense.  Unfortunately, the quoted errors on \RSUN and \VSUN are
not normal, and the values themselves are not averages in the
statistical sense, but are rather {\em consensus} values with {\em
consensus} errors.  These arguments lead to the conclusion that the {\em
values} of the Galactic constants and their errors {\em can not} be used
in analyses like maximum-likelihood estimates of a property $X$ which
depends on the Galactic constants.  Instead we urge researchers to
investigate the dependence of their results on the assumed values of the
Galactic constants and present their conclusions as functions of \RSUN
and \VSUNn.  Given the tremendous increase in computing power, such an
approach is currently more feasible than in the past.  Practicing what
we preach, we follow this approach in previous papers \cite{rOmM98cPI}
as well as in the current article.

   In Paper~I we used both the constraints set by the flaring of the
\HI\ layer as well as the boundary conditions imposed by the local
stellar kinematics to infer the halo's flattening.  The procedure
works as follows: 1) pick values for \RSUN and \VSUN and determine the
corresponding rotation curve, 2) create mass models with as
ingredients the stellar bulge and disk, the interstellar medium (ISM)
and a dark halo of varying degrees of flattening, 3) pick values for
the observationally ill-determined scale-length and mass of the
stellar disk, 4) select a value for \sgas\ and calculate model flaring
curves \cite{rpO95}, 5) select the model with flattening \qHI\ such
that the observed \HI\ flaring at $R\sim2R_0$ is reproduced [this
flattening will be independent of the distribution of the stellar
mass], 6) now vary the stellar mass (\Sstr) at \RSUN, 7) keep track of
the dark matter column density \SHkg\ that needs to be added to \Sstr\
to match \SKG, 8) work out which halo flattening \qdT\ is required to
generate \SHkg, and finally, 9) select the model that identical \qHI\
and \qdT\ values. This method is equivalent to finding the zero-point
of the $[\qHI(\Sstr)-\qdT(\Sstr)]$ function. Graphically, this process
can be represented as determining the intersection of the
$\qHI(\Sstr)$ and $\qdT(\Sstr)$ curves (see Paper I, figure 3). Note
that the \qdT(\Sstr) relation hardly depends on the disk scale-lenght.

In this manner we construct self-consistent mass models, with specific
values for \hd, \Sstr\ and $q$, for the selected combination of the
Galactic constants.  Possibly the weakest link in this procedure is
the assumed value for \sgas.  In the present paper we will remedy this
shortcoming of Paper~I by repeating the procedure outlined above for
many values of the gaseous velocity dispersion.  Picking a value for
\sgas\ in the outer Galaxy is related to the question as to the
relevance of non-thermal pressure support (due to magnetic fields and
cosmic ray energy density) of the gas layer.  We investigate these
issues in detail in section
\ref{sec:The_effects_of_a_larger_gaseous_velocity_dispersion}.

   The surface density of stars in the Solar neighbourhood provides a
significant constraint on the determination of the halo's shape.  We
therefore review recent determinations in section
\ref{sec:The_local_stellar_column_density_a_constraint}.  Kuijken \&
Gilmore (1991, henceforth KG91) claim that the columndensity within 1.1
kpc of the plane is much better determined than the values of the
individual components.  However, their Galactic constants lie at the
high end of the range we use.  Since \SKG\ provides us with such an
important constraint, we consider it prudent to check KG91's assertion. 
We indeed confirm KG91's finding that \SKG\ is relatively well determined
over a large range in Galactic constants, \Sstr\ and \hd\ (Appendix
\ref{app:appendix_Kz}).  In the following section we describe our mass
models in more detail and determine the density of dark matter in the
Solar neighbourhood.  We summarize and conclude in \S
\ref{sec:Summary_and_Conclusions}.

\mVskip{-5mm}
  \section{Mass Models}
\label{sec:Mass_Models}

In order to interpret the available data we build axisymmetric Milky
Way mass models for which we determine the model values of the total
disk mass, the \HI\ flaring, and other parameters. It is well-known
that the Milky Way deviates from azimuthal symmetry. However, as we
have shown in Paper I, and will do so again below, the current
observational constraints are only barely good enough to rule-out the
most extreme combinations of Galactic constants. Thus, it would be
unrealistic to try to incorporate fine-structure in the stellar mass
distribution due to a bar and/or spiral structure, especially since
these features are not terribly well determined themselves. This
situation may dramatically change with the advent of future
astrometric satellites (e.g., DIVA, FAME, GAIA, SIM) that could
determine the stellar column density over large parts of the Galactic
disk.

A comparison between the observations and the models yields the size,
mass and shape of the Milky Way.  Our models include: a stellar bulge
and disk, a gas layer, and a dark halo.  Some of the relevant
parameters of our models are tabulated in
Table~\ref{tab:Some_bulge_disk_and_halo_parameters}.  We consider
models with a range of Galactic constants, disk exponential
scale-length, stellar disk mass, total columndensity, and halo
flattening.  To calculate the {\em exact} vertical force law
(\Kz{(R,z)}) at every point $(R,z)$ in the Galaxy we integrate over
the full mass distribution: $\Kz{(R,z)} = G \int_0^{\infty} r dr
\rho(r,0) \int_{-\infty}^{\infty} dw \rho(r,w) \int_{-\pi}^{\pi}
\frac{d}{dz} \frac{d\theta}{|\overline{\bf s} -\overline{\bf S}|}$ with
$\overline{\bf s}=\{r,w\}, \overline{\bf S}=\{R,z\}$, $\rho$ the total
mass density at $(R,z)$, and $G$ Newton's constant of gravity
\cite{rpO95}. Since the calculation of \Kz{} and the model flaring
curve is rather expensive\footnote{The calculation of model gas layer
widths for a given combination of \RSUNn, \VSUNn, \hd\ and bulge and
disk mass-to-light ratios takes about 1.3 hours per $q$-values, on a
SPARC-10 processor.}  we perform these detailed
calculations only for the limited subset of models listed in
Table~\ref{tab:Disk_Halo_parameters}.  We determine the model flaring
curves for a much larger range in \RSUN and \VSUN values by applying
an approximation (Olling 1995, Appendix D) which we calibrate using
the models presented in Table~\ref{tab:Disk_Halo_parameters}, see
section \ref{sec:Constraints_from_the_thickness_of_the_gas_layer} for
further details.
	
\mVskip{-2mm}
\subsection{The Mass Components}
\label{sec:The_Mass_Components}

\subsubsection{Stellar Components}

We base our model for the bulge on Kent's (1992) K-band luminosity
distribution.  The bulge is a modified spheroid with ``boxy'' appearance
and density: $\rho_b(s) = d_b ~ K_0(s/h_b)$, with $s^4 = R^4 + (z/q_b)^4
+ \zeta_b^4$.  The bulge is flattened with axial ratio $q_b$.  To avoid
the singularity of the $K_0$-Bessel function at $s=0$, we include a
softening $\zeta_b$.  We also truncate the bulge exponentially beyond 3
kpc.

\begin{table}
 \caption{Some parameters used in the model calculations. }
 \label{tab:Some_bulge_disk_and_halo_parameters}
\begin{tabular}{l|r|l}
                        &          & \\ \hline
bulge:                  &          &                           \\
   $d_b$                & 3.53     & $L_{\sun,K}$ pc\rtp{-3}   \\
   $q_b$                & 0.61     &                           \\
   $h_b$                & 667      & pc                        \\
   $\zeta_b$            &  10      & pc                        \\
   $\rho_b(0)$          &  15.24   &  $L_{\sun,K}$ pc\rtp{-3}  \\
   L$_{b,K,tot}$        & 1.5 10\rtp{10}  & \LsunB{K}          \\ \hline
disk:                   &                 &             \\
   $z_e$                & 300             &  pc         \\
   $h_d$                & 2, ~2.5, ~3     & kpc         \\
   $L_{\rm d,K}(0)$     & 2200, 1408, 978 &
   $L_{\sun,K}$ pc\rtp{-2} (\hd=2,2.5,3) \\
   \Sstr(\RSUNn)        & 25-45           & \LSpcsq     \\
   $\rho_{*}$           & 26.5 - 47.7     & 10\rtp{-3} \MSpccub \\
   L$_{d,K,tot}$        & 5.5 10\rtp{10}  & \LsunB{K}   \\ \hline
ISM at \RSUNn:                             &                 &         \\
   $\Sigma_{\HI+\HII}$  & 9.25            & \MSpcsq \\
   $\Sigma_{\Ht}$       & 1.80            & \MSpcsq \\
   \FWHM{_{\HI}}        & 410 \pmt 30     & $(R_0/7.1)$ pc      \\
   \FWHM{_{\Ht}}        & 141 \pmt 20     & $(R_0/7.1)$ pc      \\
   $\rho_{\HI+\HII}$    & ~21 \pmt 1.5    & $(7.1/R_0)$ 10\rtp{-3} \MSpccub \\
   $\rho_{\Ht}$         & ~12 \pmt 1.5    & $(7.1/R_0)$ 10\rtp{-3} \MSpccub \\
   $\rho_{ISM}$         & ~43 \pmt 3.0    & $(7.1/R_0)$ 10\rtp{-3} \MSpccub \\ \hline
\end{tabular}
\end{table}

For the disk, we use the standard form of a radially exponential disk
with central K-band surface brightness $L_{\rm d,K}(0)$ and scale-length \hd.
We will consider models with values for the scale-length that are
considered to be reasonable \cite{pdS97}.  Furthermore we use the
observational fact that the total luminosity is better determined than
either the central surface brightness or the scale-length
\cite{KDF91}.  Thus, we scale $L_{\rm d,K}(0)$ such that the total
luminosity is conserved for each choice of \hd.  However, the
numerical values of $L_{\rm d,K}(0)$ and \hd\ depend on our choice of
\RSUNn.  From Freudenreich (1996, 1998) we find that these quantities
typically increase with \RSUNn, by $8\% \times$ (\RSUNn/kpc - 8).
Such variations induce a change in total luminosity which is a factor
of three larger.  However, because these corrections are smaller than
the changes resulting from the uncertainty in \hd, we do not scale
L$_{d,K,tot}$ with \RSUNn.  For the disk's vertical distribution we
use a secant-hyperbolic function [sech($\frac{z}{2*z_e}$)], which is a
compromise between the often used exponential and
secant-hyperbolic-squared forms \cite{pvdK88}.  For simplicity, the
exponential scale-height (\ze) of the disk is taken to be constant at
300 pc, intermediate between the values suggested by Kent (1992) and
Reid \& Majewski (1993).  But note that our results depend neither on
\ze\ nor on the particularities of the vertical density distribution.
We use disk mass-to-light ratios (\Ups{d,K}) such that Kuijken \&
Gilmore's (1989b) 2-$\sigma$ range for the local stellar columndensity
is spanned.  The stellar disk is truncated at (\RSUN + 4.5) kpc
\cite{RCM92,htF9698}.

\subsubsection{Interstellar Medium Components}

We now turn to the contribution to the mass models from the
interstellar medium.  Our location inside the Milky Way means that
distances to diffuse components like the ISM are based on a
kinematical model.  Thus, important properties like the full width at
half maximum (FWHM), volume density, and total mass of the ISM depend
on the Galactic constants and the Milky Way's rotation curve.  Thus,
we re-determine the atomic and molecular gas distributions for each
choice of \RSUN and \VSUN.  While the columndensities at {\em
fractional radius $R/\RSUNn$} are independent\footnote{From Binney \&
Merrifield's (1998) equation~9.12 it follows that the
distance-dependent function that appears in the exponential term only
depends on $r=R/R_0$, so that one can rewrite the observed brightness
temperature at line-of-sight velocity $u_{los}$ as $R_0 \int dr F(r,
...) n(r) H(u_{los},r, ...)$ with $F$ a function that results from the
transformation from line-of-sight distance to $r$, and $H$ the
velocity-dependent function.  If $H$ varies more quickly with $r$ than
the number density $n(r)$, then the product $R_0 \times n(r)$ is
approximately constant.  Thus, as $R_0$ is increased, the volume
density goes down.  Because the gas-layer width {\em increases} with
\RSUNn, the vertical column-density at scaled distance $R/R_0$ is {\em
independent} of the Galactic constants.  See also Bronfman \etal
(1988).} of the choice of \RSUN \& \VSUNn, the number density, the
thickness and the total gas mass of the Galaxy are not.
Observationally, we can currently do no better than determining the
gaseous column-density (\Sgas) at fractional radius $R/R_0$.
Likewise, since the thickness measurement returns an angular size, we
can only determine FWHM/\RSUN at fractional radius $R/R_0$ [FWHM$/R_0
= \zeta R/R_0$, \cite{jBmM98}, their figure~9.25].  Physical
length-scales for the thickness and Galactocentric radius can be
assigned after choosing a value for
\RSUNn.

For the inner Galaxy, the \HI\ columndensity was determined from the
midplane volume density \cite{bB88} and the observed thickness of the
layer \cite{sM95}.  The \HI\ columndensities for $R \ga R_0$ were
taken from Wouterloot \etal (1990).  The \Ht\ columndensities for the
inner and outer Galaxy were copied from Bronfman \etal (1988) and
Wouterloot \etal (1990), respectively\footnote{A graphical
representation of the derived gaseous surface density distributions
can be found in a related paper on the Galactic \& Oort constants
\cite{rOmM98c}.  The \Ht\ columndensities have been re-scaled
from the original sources assuming $N(\Ht)/W(CO) = 2.3 \, 10\rtp{20}$
cm\rtp{-2} (K \kms)\rtp{-1}. We present the radial increase of the
thickness of the gas layer in Paper~I. The thickness and columndensity
of the \HI\ and \Ht\ are tabulated in Appendix~\ref{app:ISM}}.  For
the radial profiles of the ISM we neglect the columndensity due to the
other phases of the ISM. However, in order to properly take into
account all baryonic contributions to the local columndensity, we
include 1.4 \MSpcsq\ of ionized hydrogen \cite{srKcH87} {\em at the
Solar position only}.  We further include 23.8\% helium by mass
\cite{kaOgS95} to compute the surface densities.

As in our previous papers, we do not distinguish between the various
phases of the atomic Hydrogen but rather assume that the warm neutral
medium has a kinetic temperature equivalent to the ``temperature''
associated with the bulk motions of the clumped, cold neutral
medium. Note that the cold medium is likely to be absent beyond the
stellar disk \cite{PaperI,rB97}

\subsubsection{The Dark Matter Component}

\begin{figure}
 \epsffile{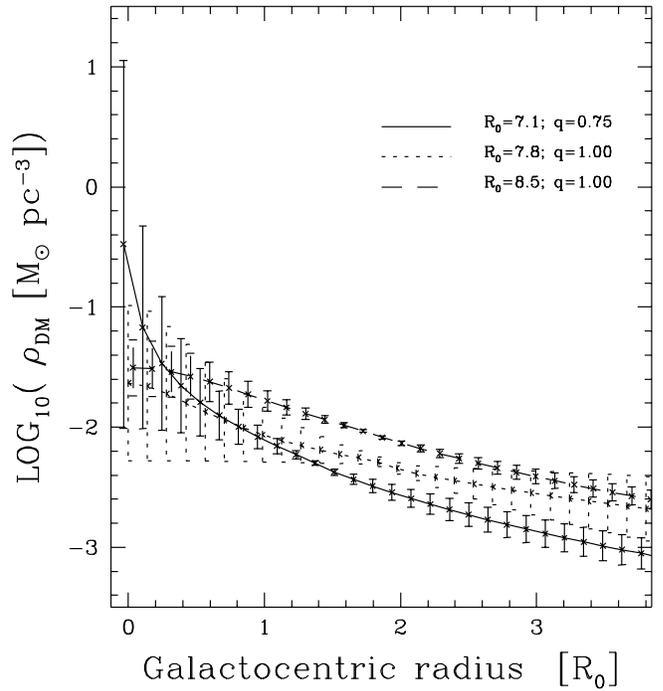}
 \caption{\label{fig:MW_DM_density_distributions} The dark matter
midplane density distribution calculated using equation 
(\protect\ref{eqn:rho_halo_Rz}) for three values of \protect\RSUN and $q$. 
The densities scale approximately as $1/q$ (Olling 1995) The error bars
represent the full range for the case corresponding to $\Sstr = 35 \pm
5$ \MSpcsq, and \hd=2 to 3 kpc.  Notice that although the DM density at
the Solar circle is relatively well determined, the central density is
uncertain by more than three orders of magnitude. Note that we have
offset the radial coordinates of the $R_0$=7.1, and 8.5 kpc
models from their true positions (i.e., the $R_0$=7.8 kpc points) to
avoid overlapping symbols.
}
\end{figure}

  As is commonly done
\cite{egvAS86,smK87,kBeg89,LF89,pdSlS90,ahB92,rpO95,rpO96b}, we model
the dark matter density distribution as a non-singular isothermal
spheroid with flattening $q$, and a density distribution given by:

\negSMLskip
\begin{eqnarray}
\rho_h(R,z;q) &=& \rho_0(q) \;
  \left( \frac{R_c^2(q)}{R_c^2(q) + R^2 + (z/q)^2} \right) \; , 
  \label{eqn:rho_halo_Rz}
\end{eqnarray}

\noindent where the halo's core radius (\Rc) and central density (\rhoN)
depend on the flattening in such a way that the family of density
distributions $\rho_h(q)$ have a rotation curve that is essentially
independent of $q$ \cite{rpO95}. 
Figure~\ref{fig:MW_DM_density_distributions} shows the radial
distribution of dark matter for three values of \RSUNn, for the range in
model parameters as listed in Table~\ref{tab:Disk_Halo_parameters}. 
Each choice for \hd\  and \Sstr\  results in a different distribution. 
The largest DM densities are obtained for large \hd's and small \Sstr's. 
Even though the central DM density is uncertain by over three orders of
magnitude, the local dark matter density is determined rather well: we
find

\begin{eqnarray}
\frac{\rho_{\rm DM}(\RSUNn,\VSUNn)}{10\rtp{-3} \MSpccub} &=&
   \frac{11.5 + 3.8 \times (\RSUN-7.8) \pm 2}
   {q \left( 26.7/\Omega_0 \right)^2}
   \label{eqn:rhoDM_R0}
\end{eqnarray}

\noindent with $\Omega_0 = \Theta_0/R_0$, and where \RSUN is in kpc, and
\VSUN in \kms.  The $(\sim25/q)$\% uncertainty arises as a result of
the uncertainty in \SKG\ and \Sstr.  For example, taking $R_0=7.1$
kpc, $q=0.71$, and using the parameters tabulated in Table 2, we find
$\rho_{DM}(R_0) = 10.5$ m\MSpccub~(1 m\MSpccub = 10\rtp{-3} \MSpccub).
Thus, the $R_0=7.1$ model value for the DM matter density compares
well with the values given by equation~(\ref{eqn:rhoDM_R0}) as well as
with the observational determination (\S
\ref{sec:The_local_stellar_column_density_a_constraint}).

\begin{table*}
\caption{Disk \& halo parameters.  A log for a representative sample of
models for which {\em exact} flaring curves are calculated.  The
meaning of the columns is as follows: 1) $\gamma$, the degree to which
the stellar disk is maximal; 2) scale-length of stellar disk (kpc);
3\&4) bulge \& disk K-band mass-to-light ratios; 5\&6) stellar
columndensity \& total columndensity within 1.1 kpc of the plane
(\protect\MSpcsq); 7\&8) core radius (kpc) and central density
(m\protect\MSpccub) of the dark halo, assuming $q$=1 (multiply by
$\approx 1/q$ to scale to different values of $q$, for a more exact
scaling, see Olling 1995); 9\&10) dark halo flattening as inferred
from the \protect\HI\ flaring (\protect\qHI) and the local
columndensities (\protect\qdT). For a given set of
(\RSUNn,\VSUNn,\hd), the \protect\qHI\ errors for the Milky Way are
similar to those for extra-galactic systems \protect\cite{rpO96a} .}
\label{tab:Disk_Halo_parameters}
\begin{tabular}{|r|r|r|r|r|r|r|r|c|c|}
        &  & &     &     &    &      &     &                  & \\ \hline \hline
$\gamma$&\hd &\Ups{b,K}&\Ups{d,K} &\Sstr &\SKG &\Rc  &$\rhoN(q=1)$ & \qHI     &      \qdT \\ \hline \hline
\RSUNn=6.8, & \VSUNn=182  & &     &     &     &     &      &                  & \\ \hline
0.80&2.0&0.55&0.422  &30.21&62.48& 2.39&  73.8& 0.630 \pmt 0.107 & 0.561 \pmt 0.175 \\
0.86&2.0&0.55&0.488  &34.92&66.02& 3.54&  35.6& 0.687 \pmt 0.113 & 0.664 \pmt 0.226 \\
0.92&2.0&0.55&0.558  &39.96&68.72& 6.09&  14.6& 0.699 \pmt 0.120 & 0.753 \pmt 0.295 \\
0.64&2.5&0.55&0.327  &29.58&63.56& 1.32& 246.3& 0.687 \pmt 0.152 & 0.620 \pmt 0.183 \\
0.70&2.5&0.55&0.392  &35.39&68.84& 1.86& 125.1& 0.717 \pmt 0.134 & 0.826 \pmt 0.268 \\
0.75&2.5&0.55&0.449  &40.63&73.40& 2.45&  73.0& 0.674 \pmt 0.127 & 1.061 \pmt 0.371 \\
0.56&3.0&0.55&0.306  &30.21&64.34& 0.70& 853.9& 0.740 \pmt 0.181 & 0.651 \pmt 0.194 \\
0.60&3.0&0.55&0.351  &34.68&68.6 & 1.00& 421.0& 0.770 \pmt 0.159 & 0.820 \pmt 0.261 \\
0.65&3.0&0.55&0.412  &40.70&74.18& 1.44& 201.9& 0.762 \pmt 0.137 & 1.114 \pmt 0.382 \\ \hline
\RSUNn=7.1, & \VSUNn=189  & &     &     &     &     &      &                  & \\ \hline
0.82&2.0&0.45&0.482  &29.71&61.94& 2.73&  62.7& 0.625 \pmt 0.132 & 0.544 \pmt 0.165 \\
0.89&2.0&0.45&0.568  &35.00&65.66& 4.53&  25.5& 0.624 \pmt 0.116 & 0.636 \pmt 0.214 \\
0.95&2.0&0.45&0.647  &39.87&66.74& 9.77&   8.6& 0.679 \pmt 0.104 & 0.603 \pmt 0.239 \\
0.65&2.5&0.55&0.370  &29.66&62.60& 1.13& 340.1& 0.727 \pmt 0.159 & 0.581 \pmt 0.173 \\
0.71&2.5&0.55&0.441  &35.39&67.76& 1.66& 158.2& 0.738 \pmt 0.161 & 0.765 \pmt 0.250 \\
0.76&2.5&0.55&0.506  &40.55&72.26& 2.25&  86.4& 0.823 \pmt 0.223 & 0.999 \pmt 0.356 \\
0.56&3.0&0.55&0.333  &29.73&61.82& 0.20&9884.5& 0.735 \pmt 0.181 & 0.550 \pmt 0.168 \\
0.61&3.0&0.55&0.395  &35.28&67.10& 0.55&1343.6& 0.703 \pmt 0.158 & 0.734 \pmt 0.243 \\
0.63&3.0&0.55&0.421  &37.63&69.32& 0.71& 802.8& 0.786 \pmt 0.210 & 0.843 \pmt 0.290 \\ \hline
\RSUNn=7.8, & \VSUNn=207 & &     &     &     &     &      &                  & \\ \hline
0.92&2.0&0.45&0.687  &29.80&64.60& 9.51&  15.4& 1.003 \pmt 0.119 & 0.616 \pmt 0.162 \\
0.72&2.5&0.75&0.492  &29.80&64.76& 4.59&  47.9& 1.744 \pmt 0.553 & 0.963 \pmt 0.240 \\
0.78&2.5&0.75&0.577  &34.97&71.00& 5.81&  32.6& 1.616 \pmt 0.443 & 1.065 \pmt 0.260 \\
0.84&2.5&0.75&0.669  &40.56&77.66& 8.01&  20.6& 1.461 \pmt 0.337 & 1.191 \pmt 0.331 \\
0.59&3.0&0.75&0.430  &30.46&66.54& 2.84& 103.4& 1.623 \pmt 0.451 & 0.973 \pmt 0.244 \\
0.61&3.0&0.75&0.460  &32.56&68.06& 3.02&  91.7& 2.019 \pmt 0.758 & 1.026 \pmt 0.260 \\
0.68&3.0&0.75&0.572  &40.46&77.54& 3.81&  59.3& 1.810 \pmt 0.753 & 1.327 \pmt 0.364 \\ \hline
\RSUNn=8.5, & \VSUNn=227  & &     &     &     &     &      &                  & \\ \hline
0.76&2.5&0.60&0.670  &30.66&78.68& 6.81&  39.1& 2.922 \pmt 0.863 & 1.169 \pmt 0.252 \\
0.81&2.5&0.60&0.761  &34.83&80.43& 8.34&  28.9& 2.512 \pmt 0.608 & 1.228 \pmt 0.279 \\ 
0.86&2.5&0.60&0.857  &39.26&81.28&11.10&  19.9& 1.931 \pmt 0.366 & 1.256 \pmt 0.306 \\
0.59&3.0&0.85&0.486  &27.23&76.53& 5.09&  60.2& 1.424 \pmt 0.156 & 1.165 \pmt 0.193 \\
0.67&3.0&0.85&0.627  &35.11&82.49& 6.20&  43.2& 1.498 \pmt 0.183 & 1.364 \pmt 0.213 \\
0.75&3.0&0.85&0.785  &44.00&88.53& 8.07&  28.9& 1.475 \pmt 0.197 & 1.641 \pmt 0.159 \\ \hline
\end{tabular}
\end{table*}

\mVskip{-2mm}
\subsection{Other Dark Matter models}
 \label{sec:Other_dark_matter_models}

Of course, other mass models can be constructed which represent the
radial dark matter density distribution \cite{egvAS86,NFW96,wDjB98}.
However, all viable mass models must share the property that they
reproduce the Galactic rotation curve.  For a round halo the vertical
force approximately equals $z/R$ times the radial force.  Since the
radial force is approximately the same for all models which reproduce
the observed rotation curve, the ensemble of possible models also have
approximately equal vertical forces, independent of the exact radial
DM density distribution \cite{rpO95}.  In a flattened halo with the
same rotation curve, the DM densities are roughly proportional to
$1/q$, independent of the radial mass distribution [cf.
eqn.~(\ref{eqn:rhoDM_R0})].  Thus, our analysis will not be seriously
compromised by restricting ourselves to one particular DM density
distribution.

\mVskip{-5mm}
  \section{Constraining the Dark Matter Density}
\label{sec:Constraining_the_Dark_Matter_Density}

In this section we present two simple analytical models to illustrate
the existing correlations between \RSUNn, \VSUNn, \SKG, \Sstr, $q$, and
\sgas\  outlined in the Introduction.  These models can be used to show
how the local stellar kinematics and the \HI\  flaring constrain the dark
matter density in the Solar neighbourhood and at $\sim2$ \RSUNn.  But
first we investigate how well the local stellar columndensity is
determined and how accurately this determination constrains the dark
matter density in the Solar neighbourhood.

\mVskip{-2mm}
\subsection{The local stellar columndensity: a constraint?}
 \label{sec:The_local_stellar_column_density_a_constraint}

The Milky Way is a unique galaxy in that we can, at least in principle,
determine the local columndensity of stars directly.  Once \Sstr\ is
accurately determined, it is possible to establish to what degree the
Milky Way disk is maximal, which would provide an important benchmark
for external galaxies.  Unfortunately this benchmark is not yet
available, as the values for \Sstr\ reported in the literature range
from 26 to 145 \MSpcsq. 

Two basic techniques have been employed to determine \Sstr.  The
direct method involves converting star counts as a function of
Galactic coordinates and magnitude to in-situ mass densities.  This
method is somewhat hampered by uncertainties in the conversion from
luminosity to mass, completeness problems in the Solar neighborhood,
and binary corrections at large distances.  Gould, Bahcall \& Flynn
(1997; hereafter referred to as GBF97) used deep HST star counts of
M-dwarfs at great heights above the plane in combination with a local
normalization to infer a stellar columndensity of only 25.8 \pmt 3.8
\MSpcsq.  The second, kinematical, method employs the
interrelationship between the potential, the vertical density
distribution, and the variation of the velocity dispersion with $z$.
Many authors have employed this method, yielding a large range in
inferred values for \Sstr.  For example, Bahcall (1984b) found that
the inferred value of \Sstr\ depends significantly on the assumed
vertical distribution of the dark matter: $\sim$40, $\sim$52,
$\sim$68, and $\sim$145 \MSpcsq\ for dark matter distributions that
resemble the gaseous disk, an isothermal halo, the thin stellar disk,
or the thick stellar disk, respectively.  Other authors find values as
low as 35 \MSpcsq\ \cite{kKgG89b,cFbF94}.  Methods that combine the
two primary methods exist as well \cite{BRC87,CRB89}.  In Appendix
\ref{app:appendix_Kz} we simulate the kinematic determination of the
stellar columndensity with the aid of Galaxy models --taken from
Table~\ref{tab:Disk_Halo_parameters}-- for which we know the exact
force law.  Our results are in complete agreement with the findings of
previous authors: the ``vertical disk-halo conspiracy'' can only be
resolved if high-$z$ stellar kinematical data are included and/or if
additional assumptions are made (e.g., a ``reasonable'' value for the
DM density).  Typical values for the mass of the {\em stellar} disk
are 52 \cite{jnB84b}, 35 \pmt 5 \cite{kKgG89b} and 37 \pmt 13 \MSpcsq
\cite{cFbF94}.

Considering the uncertainties in the kinematical estimates of the total
mass of the disk, it might be preferable to use the direct star count
method to determine \Sstr.  The latest results by GBF97 imply \Sstr=25.8
\pmt 3.8 \MSpcsq.  However, the local space density of stars found by
these authors [$(33.2 \pm 8.6$) m\MSpccub] is somewhat lower than the
values reported in the literature [(43 \pmt 15)\footnote{The average
of: $\rho_*$ = 46 \cite{rW74}, 50.8 \cite{jnB84a,BFG92}, 45 m\MSpccub\
[Cr\'{e}z\'{e} \etal (1998).} m\MSpccub].  This suggests that GBF97
may have under-estimated \Sstr\ by a factor of 1.4 \pmt 0.2, and so we
find \Sstr\ = 36 \pmt 5 \MSpcsq: the star count method yields a value
for \Sstr\ that is remarkably close to the kinematical estimates.

On the other hand, the total matter density in the Solar neighbourhood
as inferred from recent Hipparcos data \cite{CCBP98} of 76 \pmt 15
m\MSpccub\ favours a local stellar volume density which is even smaller
than reported by GBF97.  After subtracting the density of the ISM (see
Table~\ref{tab:Some_bulge_disk_and_halo_parameters}) and the local DM
density [eqn.~(\ref{eqn:rhoDM_R0})] we find a local stellar density of
$\sim 21 \pm 15$ m\MSpccub.  Note that even lower values for $\rho_*$
result if large values for the Galactic constants and/or $\Omega_0$ are
chosen, while smallish Galactic constants and $\Omega_0$ {\em increase}
the local stellar density. 

To summarize, a ``reasonable'' value for \Sstr\ might be 35 \MSpcsq,
and we suggest a ``consensus'' error of 10 \MSpcsq.  However, since
the differences between the various \Sstr\ estimates are not random
but systematic, one {\em should not} interpret the reasonable \Sstr\
value and its error in a statistical sense. That is to say, {\em one
can not construct likelihood contours based on a reasonable average
value and a consensus error bar.} Any attempt to do so would be an
over-interpretation of the available data.

A ``reasonable'' value for the local dark matter column can be
obtained by subtracting the columndensities of the stellar and ISM
distributions: $\Sigma_h$ = (24.5 \pmt 11) \MSpcsq\ within 1.1 kpc of
the plane. This dark matter column amounts to an average local dark
matter density of 11 \pmt 5 m\MSpccub.

\mVskip{-2mm}
\subsection{The connection between dark and luminous matter in the Solar neighbourhood}
 \label{sec:The_connection_between_dark_and_luminous_matter}

In the previous section we have seen that the contribution of the
stellar and dark matter components to the local disk mass cannot be
clearly segregated.  Thus the observed value for \SKG\  implies that
\Sstr\  and $\Sigma_h$ are highly correlated: a low stellar column
implies a large amount of dark matter, and vice versa.  Since the dark
matter density depends on the amplitude of the rotation curve and the
halo's flattening, these parameters are in turn related to \Sstr.  Below
we investigate the relations between \SKG, \VSUNn, \Sstr\  and $q$ in
some detail. These relations are independent of the \HI\ flaring.

Integrating equation (\ref{eqn:rho_halo_Rz}) with respect to $z$ and
applying equation (A4) from Olling (1995), we find the columndensity of
dark matter within $z$ of the Galactic plane as a function of $q$:

\negSMLskip
\begin{eqnarray}
\Sigma_h^z(q) \hspace*{-1ex} &=& \hspace*{-1ex}
   \frac{2 q \rho_0(q) R_c^2(q)}{\sqrt{R_c^2(q) + R^2}}
   \arctan{\left(\frac{z/q}{\sqrt{R_c^2(q) + R^2}}\right)}
   \label{eqn:Sigma_DM1p1} \\*[3mm]
   &&\hspace*{-1.6cm} = 
   \frac{V_{h,\infty}^2 \sqrt{1-q^2}}{2 \pi G  \sqrt{R_c^2(q) + R^2} \arccos{q}}
   \arctan{\left(\frac{z/q}{\sqrt{R_c^2(q) + R^2}}\right)} \, ,
   \label{eqn:Sigma_DM1p1_Vhalo}
\end{eqnarray}

\noindent with $V_{h,\infty}$ the asymptotic rotation velocity of the
round dark halo.  For each of our self-consistent mass models we can
calculate $\SKG(q)$ and find that it obeys a simple power law
relation:

\negSMLskip
\begin{eqnarray}
\SKG(q) \hspace*{-2mm} &=& \SHkg(q) + \Sstr + \Sgas
   \label{eqn:SKG_his} \\
                 &\approx&  \hspace*{-2mm}
   \SKG(q=1) \times q\rtp{-p} \, .
   \label{eqn:SKG_q}
\end{eqnarray}

\noindent The dependence of \SKG\ on the other parameters of the model
parameters are combined into the index $p$ (with $p \sim0.05 -
0.45$)\footnote{The values for $\SKG(q), ~\SKG(q=1)$ and $p$ are
different for each combination of the Galactic constants, the stellar
disk mass and scale-length.  $p$ can be determined from a fit to
equation~(\ref{eqn:SKG_q}), where we determine $\SKG(q)$ for several
models with varying $q$'s.  Note that we also use
equation~(\ref{eqn:SKG_q}) to extrapolate into the prolate regime,
with the exponent $p$ as determined for the oblate models.}.  The
halo flattening inferred from the local stellar kinematics (\qdT) is
then found by equating $\SKG(q)$ with the observed value and solving
equation~(\ref{eqn:SKG_q}) for $q$

\begin{eqnarray}
\qdT &=& \left( \frac{\SKG(q=1)}{71 \pm 6} \right)^{\frac {1}{p}} \, .
   \label{eqn:q_dT}
\end{eqnarray}

\nid Equations~(\ref{eqn:Sigma_DM1p1_Vhalo}), (\ref{eqn:SKG_q}) and
(\ref{eqn:q_dT}) can be combined to find the dependence of \qdT\ on
the Galactic constants. Neglecting \Rc\ and equating $V_{h,\infty}$
with
\VSUNn, we find:

\begin{eqnarray}
\qdT &\propto& 
   \left( \frac{c \Theta^2_0/R^2_0 + \Sstr + \Sgas}{\SKG} \right)^{1/p} \, 
   \label{eqn:qdT_T0_R0} \, ,
\end{eqnarray}

\nid where $c$ is a constant. Qualitatively, this is the functional
dependence we expect: increasing \VSUN at constant \RSUN would
increase the required amount of DM, which has to be counteracted by
increasing \qdT\ to retain the same \SKG. Likewise, increasing \RSUN
would place the Sun in a lower density region of the halo, which needs
to be compensated by decreasing \qdT. Furthermore, given the small
values of $p$, \qdT\ depends very strongly upon the Galactic
constants.

In Appendix~\ref{app:appendix_q_Sigma_Theta} we present some other
correlations between the Galactic constants, \Sstr, \VSUN and \qdT.
For example, equation~(\ref{eqn:q_ga}) reveals the linear relation
between \qdT\ and \Sstr\ for $q\ga 0.15$. Note that models with
different scale-lenghts follow the same \qdT(\Sstr) relation, albeit
that different \hd's yield different \qdT's for a given stellar column
density\footnote{The reader can verify this by plotting the
\qdT\ values as a function of \Sstr\ from
table~\ref{tab:Disk_Halo_parameters}.}.  These examples show that the
\SKG\ constraint implies highly flattened DM halos for small values of
\VSUNn.

\mVskip{-2mm}
\subsection{Constraints from the thickness of the gas layer}
 \label{sec:Constraints_from_the_thickness_of_the_gas_layer}

In an idealized picture, the equilibrium thickness of the gas in the
Milky Way depends only on the gas ``temperature'' and the form of the
potential in which it has settled.  Thus, the observed FWHM of the gas
layer can be used to constrain the potential of the Galaxy.  In
Paper~I we present evidence that the interstellar medium beyond the
optical disk comprises only a single, iso-thermal component.  We
therefore adopt the same assumption in the analysis below.  In this
section we will expand our analysis to include the effects of
non-thermal pressure gradients.

If the gas layer is in a steady state and we assume that only
``thermal'' motions of the gas contribute to the pressure in the ISM,
the thickness of the gas layer follows from the equation of hydrostatic
equilibrium:

\negSMLskip
\begin{eqnarray}
\frac{d \ \sgas^2 \rho_g(z)}{d z} &=& \rho_{\rm g}(z) \ \Kz(z) \ ,
   \label{eqn:Hydrostatic_Equilibrium}
\end{eqnarray}

\noindent where \sgas, $\rho_g(z)$, and \Kz\ are the gaseous velocity
dispersion and volume density, and the vertical force per unit mass,
respectively.  The vertical force is commonly determined using a
``local approximation'' which \Kz\ follows from the the Poisson
equation and the local mass densities and the radial gradient of the
rotation curve.  However, this approach is known to fail in regions
where either the mass densities or the rotation curve, or both, have
steep gradients.  Further, the local approach neglects any variation
of the circular speed with height above the plane.  These problems can
be overcome by calculating \Kz\ from the global mass distribution (the
``global approach,'' Olling 1995).  As mentioned in
section~\ref{sec:Mass_Models}, we combine the exactness of the global
approach with the speed of the local approximation for optimal
results.

If we employ the local approximation the dependencies between the
various parameters of the model become apparent.  For example, when the
potential is dominated by mass component $i$,
equation~(\ref{eqn:Hydrostatic_Equilibrium}) can be solved for the
thickness of the gas layer:

\negSMLskip
\begin{eqnarray}
\FWHM{_i} \propto \frac{\sgas}{ \sqrt{4 \pi G \rho_i}} \, ,
   \label{eqn:FWHM_rho}
\end{eqnarray}

\noindent where the proportionality constant depends on the vertical
density distribution of the dominant contributor to the local vertical
potential. We copy some relations from Olling (1995). In case the gas
is fully self-gravitating, the width is given by:

\negSMLskip
\begin{eqnarray}
\FWHM{_g} &\approx& 0.158 
   \frac{\sgas^2}{\Sgas} \, .
   \label{eqn:self_grav}
\end{eqnarray}

\nid If the potential is dominated by an iso-thermal stellar disk with
sech\rtp{2} scale-height $z_0~(z_0 = 2 z_e)$, the thickness of a gas
layer would be:

\negSMLskip
\begin{eqnarray}
\FWHM{_*} &\approx& 0.51 \sgas_{9.2} \sqrt{\frac{z_{0,0.6}}{\Sstr_{35}}} \, .
   \label{eqn:in_stellar_disk}
\end{eqnarray}

\nid or 510 parsec at the Solar circle (\sgas, $z_0$ and \Sstr\ are
expressed in units of 9.2 \kms, 0.6 kpc and 35 \MSpcsq; see also, van
der Kruit 1988). And if the dark matter halo dominates the potential,
we find:

\negSMLskip
\begin{eqnarray}
\FWHM{_h} &\approx& \sqrt{\frac{13.5 \ q}{1.4 +q}}
   \frac{\sgas}{V_{\rm h,\infty}} \sqrt{R_{c,1}^2 + R^2} \, ,
   \label{eqn:in_dark_halo}
\end{eqnarray}

\nid where $R_{c,1}$ is the core radius of the equivalent round halo
\cite{rpO95}.  All distances and widths are in kpc, and all velocities
in \kms.  The thickness of the gas layer in the combined potential of
several mass components can be solved analytically in the form of an
integral equation, but requires an iterative solution procedure
(Olling 1995, Appendix C). However, a further approximation is
possible.  Following Olling (1995, Appendix D) we use:

\negSMLskip
\begin{eqnarray}
\frac{1}{\FWHMrp{_g}{^2}} &\approx& \sum_i \frac{w_i}{\FWHMrp{_i}{^2}} \, ,
   \label{eqn:FWHM_tot}
\end{eqnarray}

\nid where the weighting factors $w_i$ reflect the relative importance
of component $i$ in the region where the gas
resides\footnote{\label{foot:w_factors} See Olling (1995) for details.
For the Milky Way we use: $w_g = ( (1.24 \Sgas + 1.55 W_h\times(\rho_h
+ \rho_r) )/(\Sgas + W_h\times(\rho_h + \rho_r))$, $w_h = ( 1.0 + 0.35
\times \exp{(-1.7W_h\rho_h/\Sgas)} )$, and $w_r = (1+w_h
)/2$. $\rho_h$ is the local dark matter density, the ``rotational''
density, $\rho_r (\equiv \rhoROT{_{rot}})$, arises due to the slope of
the rotation curve. $W_h$ equals \FWHM{_h}/2.35. For example, at
R$\sim2\RSUN, \Sgas\sim7.7$ and
$(\Sigma_h,\Sigma_r)=(3.6,-0.30),(5.3,-0.17),(6.0,-0.37)$ \MSpcsq\ for
\RSUNn=7.1, 7.8, and 8.5 kpc, respectively. These values lead to
weights that are almost independent of \RSUNn: $w_g\sim1.4,
w_h\sim1.2, w_r\sim1.1$. These weightings embody the calibration of
the local approximation for the outer Milky Way.}.  Solving
equation~\ref{eqn:FWHM_tot}, for \qHI, taking $V_{h,\infty} \approx
\VSUN$ and neglecting $R_{c,1}$, we find:

\begin{eqnarray}
&& \hspace*{-6mm} \qHI(2R_0) \approx \nonumber \\
&&   \frac{1.4} {2.4 \left( \frac{1}{(\zeta 2 R_0)^2} - 
               \left( \frac{\Sgas w_g}{0.159 \sgas^2} \right)^2
        \left( \frac{2R_0}{w_h} \frac{\sgas}{\VSUNn}\right)^2 \right) - 1} \, ,
   \label{eqn:q_HI_tot}
\end{eqnarray}

\nid with $\zeta(R) = \FWHM{_{obs}}/R$ ($\sim0.07$ for the outer Milky
Way), and where $\Sgas, w_g,$ and $w_h$ have to be evaluated at $R=2
R_0$.  From this equation we infer that the contribution of the
self-gravity of the gas depends strongly on the gaseous velocity
dispersion: low \sgas\ values result in a more negative self-gravity
contribution to the denominator, and hence leads to a larger inferred
\qHI.  As we have already mentioned, the self-gravity of the gas is an
important component as it reduces the width of the gas layer by
approximately 45\%. A further simplification can be made by neglecting
the gaseous self-gravity as well, we find:
\begin{eqnarray}
\qHI &\approx& \frac{ 1.4 [ \zeta(R) \VSUNn ]^2 }
                 { 13.5 \sgas^2 - [\zeta(R) \VSUNn]^2 } \, ,
   \label{eqn:q_zeta_Theta_sigmaG} \\
\qHI &\propto& \frac{\VSUNn^2}{\sgas^2} \, ,
   \label{eqn:qHI_T0_R0}
\end{eqnarray}

\nid where the second line arises because the first term in the
denominator dominates.  The errors in the halo flattening and rotation
speed are related through equation~(\ref{eqn:qHI_T0_errors}), where we
have neglected the contribution of the velocity dispersion error. The
dependence of \qHI\ on \VSUN and \sgas\ are indeed as outlined in the
Introduction. Also note that, due to the quadratic nature of the
proportionalities, the observed flaring and the small allowed range of
\qHI\ constrain \VSUN and \sgas\ rather tightly.

Comparing equations (\ref{eqn:qdT_T0_R0}) and (\ref{eqn:qHI_T0_R0}) we
see that the constraints on the halo shape arising from the local
stellar kinematics and the \HI\ flaring the have rather different
\RSUN and \VSUN dependencies.  Thus, it is indeed possible to learn
more about the Galactic dark matter distribution by combining these
two constraints.  Further, unlike the stellar kinematical method, the
constraints from the \HI\ flaring is independent of \Sstr\ since it
arises at $R\sim2\RSUN$, beyond the truncation of the stellar disk.

Equations~(\ref{eqn:q_HI_tot})-(\ref{eqn:qHI_T0_R0}) {\em only} serve to
illustrate the dependence of \qHI\ on the model parameters since several
important aspects are treated too simplistic.  First, neglecting $R_c$
and equating $V_{h,\infty}$ with \VSUN is only seldomly warranted. 
Second, depending on the slope of the rotation curve, an error of order
\pmt20\% is made in the inferred \qHI.  We therefore do not recommend
using equations~(\ref{eqn:q_HI_tot})-(\ref{eqn:qHI_T0_R0}) ``as is'' to
determine \qHI\ directly.  In all our calculations we employ
equation~(\ref{eqn:FWHM_tot}) where we take all mass components fully
into account, without any further simplifications.  For many
combinations of model parameters, a round halo is too dense to explain
the observed flaring (cf.  Fig.~\ref{fig:Sstr_q_R0_T0_siggas}).  Albeit
not entirely correct, we will determine the halo's contribution to the
potential using equation (\ref{eqn:in_dark_halo}) in those cases.

\mVskip{-5mm}
  \section{The effects of a larger gaseous velocity dispersion}
\label{sec:The_effects_of_a_larger_gaseous_velocity_dispersion}

Equations~(\ref{eqn:Hydrostatic_Equilibrium})-(\ref{eqn:q_HI_tot})
show that \qHI\ is a function of \sgas, \VSUN and \RSUNn.  On the
other hand, the halo flattening inferred from the local stellar
kinematics is a function of the Galactic constants and \Sstr. In a
self-consistent model, $\qdT \equiv \qHI$, so that strong
inter-relations are imposed among the currently ill-determined values
of \RSUNn, \VSUNn, \Sstr\ and \sgas.

Below we investigate how the inferred halo flattening and stellar
columndensity of a self-consistent model depend upon our choice of
\sgas.  The value of \sgas\ can have significant effects on the inferred
values of $q$ and \Sstr.  For example, in Paper~I we assumed that the
true velocity dispersion equals 9.2~\kms, and found an upper limit to
the local rotation speed of about 190 \kms.  Further, models with the
IAU recommended Galactic constants have prolate dark halos ($q \sim1.9$)
and require a rather high local stellar columndensity of $\sim$55
\MSpcsq.  Increasing \sgas\ would bring $q$ and \Sstr\ down to more
acceptable levels, and it might thus be worthwhile to treat \sgas\ as a
free rather than as a fixed parameter.

However, observations of external galaxies imply that the velocity
dispersion {\em declines} slightly in the radial range over which the
Milky Way's flaring has been measured, by a factor 1.12 \pmt 0.12
\cite{SvdK84,DHH90,jK93,sC95,rpO96b,fS97}.  Furthermore, there is some
evidence that \sgas\ in the outer Galaxy equals the value inside the
Solar circle \cite{lBdS91}.  Also, one would expect that a change in
gaseous velocity dispersion would be reflected in a change of the
residual motions of young stars with respect to the mean streaming
field.  Such changes have not been observed, neither in B stars nor in
Cepheids \cite{jBlB93,PQBM97}.  Thus, an increase in gaseous velocity
dispersion beyond the Solar circle is not likely.

\mVskip{-2mm}
\subsection{Non-thermal pressure terms}
 \label{sec:Non-thermal_pressure_terms}

In Paper~I we argued that the ISM beyond the optical disk comprises a
single, iso-thermal component (the warm neutral medium).  We now
investigate the possibility that non-thermal pressures have to be
included in the hydrostatic balance. 

It is estimated that in the Solar neighbourhood thermal motions, cosmic
rays, and magnetic fields contribute about equally to the interstellar
pressure \cite{lS78,srKcH87}.  However, note that the gas-layer width
depends on the pressure {\em gradient}, not just the pressure. 
Incorporating the non-thermal pressure terms, the equation of
hydrostatic equilibrium can be written as:

\negSMLskip
\begin{eqnarray}
\rho_{\rm g}(z) \ \Kz(z) &=&
\frac{dP_G}{d z} + \frac{dP_B}{dz} + \frac{dP_C}{dz} = \frac{dP_{tot}}{dz}
   \label{eqn:Hydrostatic_Equilibrium_non_thermal}
\end{eqnarray}

\noindent with $P_G = \sgas^2 \rho_g(z)$ the thermal gas pressure, $P_B
= B^2/8\pi$ the magnetic field pressure, and $P_C=1/3 U_C$ is the
pressure due to cosmic rays with energy density $U_C$.  Taking the
scale-heights of the kinetic, magnetic and cosmic ray energy density to
be $z_G, z_B$, and $z_C$, we write the pressure gradients as
$\frac{dP_G}{dz} =\frac{P_G}{z_G}$, $\frac{dP_B}{dz} =\frac{P_B}{z_B}$,
and $\frac{dP_C}{dz} =\frac{P_C}{z_C}$.

\subsection{Gas-Layer Support at $R_0$}
 \label{sec:Gas-Layer_Support_at_R_0}

In the Solar neighborhood the scale-heights $z_B$ and $z_C$ are
$\kappa_B$ and $\kappa_C$ times larger than the gaseous scale-height
[$\kappa_B \sim 10, \kappa_C \sim 4$; \cite{eg,srKcH87,mpR91}].
Following Spitzer (1978) we assume that the non-thermal pressure terms are
proportional to the kinetic pressure: $P_B = \alpha_B P_G$ and $P_C =
\alpha_C P_G$, with $\alpha_B \sim 0.25$ and $\alpha_C \sim 0.4$ in
the Solar neighborhood.  Using these parameterizations we find:

\begin{eqnarray}
\rho_{\rm g}(z) \ \Kz(z) &\approx& \sgas^2 
   \left(1 + \frac{\alpha_B}{\kappa_B} + \frac{\alpha_C}{\kappa_C} \right)
   \frac{d\rho_G}{dz} 
   \label{eqn:Hydrostatic_Equilibrium_sigma_gbc} \\
&\approx& (\sgasp)^2 \frac{d\rho_G}{dz} 
   \label{eqn:Hydrostatic_Equilibrium_sigma_p}
\end{eqnarray}

\nid where we have lumped all terms contributing to the hydrostatic
balance into a single unknown, the effective velocity dispersion
\sgasp. With the above simplifications and
equations~(\ref{eqn:FWHM_rho}) and (\ref{eqn:q_zeta_Theta_sigmaG}), it
becomes possible to estimate the effects of non-thermal pressure
support on the thickness of the gas layer and the inferred shape of
the dark matter halo:

\begin{eqnarray}
\left. \frac{\FWHM{_{GBC}}}{\FWHM{_{G}}} \right|_{R_0} \hspace*{-2mm}
   &\approx&  \hspace*{-2mm} \frac{\sgasp}{\sgas} =
      \sqrt{1 + \frac{\alpha_B}{\kappa_B} + \frac{\alpha_C}{\kappa_C}} \,
      \sim1.06
   \label{eqn:FWHM_non_thermal_R0} \\
\left. \frac{q_{GBC}}{q_G} \right|_{R_0}  \hspace*{-2mm}
   &\approx&   \hspace*{-2mm} \left( \frac{\sgas}{\sgasp} \right)^2 
      \sim0.89 \, ,
   \label{eqn:q_non_thermal_R0}
\end{eqnarray}

\noindent where the subscripts $GBC$ indicates that the gaseous,
magnetic and cosmic ray terms are taken into account in the
hydrostatic balance of the gas layer.  Thus, neglecting the
non-thermal contribution to the pressure balance in the local ISM
leads to a gas layer width that is under-estimated by a few percent
only.  Similarly, a slightly more flattened halo is required if
non-thermal pressure support were important in the Solar
neighborhood. 

\subsubsection{Non-Thermal Support, or Not?}

The above statements are at odds with the current paradigm [cf.
Binney \& Merrifield (1998), problem 9.7] which states that there
exists significant non-thermal pressure support of the gas layer in
the Solar neighborhood.  The argument in support of this paradigm is
that models without non-thermal pressure support under-predict the
observed gas-layer widths.  However, the model predictions depend
sensitively on the values of the Galactic constants.  Our models with
small values for \RSUN and \VSUN show only a small discrepancy between
the model and observed widths, while a large width difference exists
for models with IAU-standard Galactic constants (Paper~I, figure~5).
Furthermore, the gas-layer width depends sensitively on the local
stellar column density as well as the functional form of the vertical
density distribution. For example, the gas layer is almost 40\%
thinner in case the stellar vertical density distribution is
exponential rather than sech$^2$ \cite{pvdK88}.

In fact, if the potential in the Solar neighborhood were fully
determined by the stellar disk, and the \HI\ were truly iso-thermal,
then the predicted width via equation~\ref{eqn:in_stellar_disk} {\em
over-predicts} the observed width (cf.
table~\ref{tab:Some_bulge_disk_and_halo_parameters}).  The other
contributors to the potential decrease the model width slightly
(cf. eqn.~\ref{eqn:FWHM_tot}), in better agreement with the observed
width. We conclude that the paradigm of significant non-thermal
pressure support in the Solar neighborhood is inaccurate and that the
observed \HI\ width at the Solar circle is consistent with ``thermal''
pressure support only (cf. equation~\ref{eqn:FWHM_non_thermal_R0}).

\subsection{Gas-Layer Support at $2R_0$}
 \label{sec:Gas-Layer_Support_at_2R_0}

Since we only employ the \HI\ thickness measurements at large radii as a
constraint in our mass models, let us try to guess the value of \sgasp\
at $\sim2 R_0$. 

The gas density in the outer Galaxy is about four times smaller than at
\RSUN as a result of the decrease in columndensity (factor 2) and the
increase in thickness (factor 2).  To make a conservative estimate as to
the importance of the non-thermal terms at large distances, we assume
that the magnetic field strength, the cosmic ray energy density and
their vertical gradients equal the values in the Solar neighbourhood. 
Furthermore, we assume that \sgas\ remains unchanged.  With these
assumptions, the $\alpha$ values at 2\RSUN are four times larger than at
$R=R_0$, while the $\kappa$ values are decreased by a factor two.  Thus,
the effective velocity dispersion and the inferred halo shape are
strongly affected:

\begin{eqnarray}
\left. \frac{\sgasp}{\sgas}\right|_{2R_0} &=&
      \sqrt{1 + \frac{4\alpha_B(R_0)}{\kappa_B(R_0)/2} + 
                \frac{4\alpha_C(R_0) }{\kappa_C(R_0)/2}} \,
      \sim \, \sqrt{2}
   \label{eqn:sigrat_non_thermal_tR0_cons} \\
\left. \frac{q_{GBC}}{q_G} \right|_{2R_0} 
   &\approx& \left( \frac{\sgas}{\sgasp} \right)^2  \, \sim0.5 \, .
\end{eqnarray}

\noindent On the basis of this over-estimation, it appears that the
gas layer is significantly out of thermal equilibrium at large radii.
Do we expect this to be the case? Two effect reduce the importance of
the non-thermal contribution.  First, the B-field is probably
co-spatial, ``frozen in'', with the ionized part of the ISM which'
scale-height has most likely doubled at $R=2R_0$ (like the \HI).
Thus, the Solar neighbourhood value for $\kappa_B(2R_0)$ should be
used in equation~(\ref{eqn:sigrat_non_thermal_tR0_cons}).  Second,
radio-continuum measurements of external galaxies suggest that high
cosmic ray fluxes are closely associated with the sites of star
formation \cite{mdBgH90}.  Since active star-formation ceases to exist
beyond $\sim1.5 R_0$, the $\alpha_C$-term in
eqn.~(\ref{eqn:sigrat_non_thermal_tR0_cons}) vanishes.  Thus, in a
more realistic treatment of the non-thermal pressure terms, their
effect on the vertical equilibrium of gas at large radii is much
reduced:

\begin{eqnarray}
\left. \frac{\sgasp}{\sgas}\right|_{2R_0} &=&
      \sqrt{1 + \frac{4\alpha_B(R_0)}{\kappa_B(R_0)} } \,
      = \, \sqrt{1.1} \, \sim1.05
   \label{eqn:sigrat_non_thermal_tR0_real} \\
\left. \frac{q_{GBC}}{q_G} \right|_{2R_0} 
   &\approx& \left( \frac{\sgas}{\sgasp} \right)^2  \, \sim0.9 \, .
\end{eqnarray}

\noindent Further, if the magnetic field strength decreases with
Galactocentric radius, the above estimates should be even closer to
unity.  Thus realistic estimates of the importance of non-thermal
pressure terms in the equation of hydrostatic equilibrium indicate that
the the thickness of the \HI\  layer is not much affected, neither in the
inner, nor in the outer Galaxy. 

And finally it is worth mentioning that there exist both theoretical and
empirical evidence that non-thermal pressure support is small even when
simplified calculations indicate that they may dominate $dP_G/dz$.  From a
theoretical perspective one finds that the vertical balance is only
affected if the magnetic field is horizontally stratified, a
configuration that is unstable \cite{enP66}, and which is thus not
expected to exist on a global scale in the Milky Way.  In fact, 3D
simulations of the growth of the Parker instability show that the gas is
almost entirely supported by thermal pressure within the first four
scale-heights \cite{KHRJ98}.  On the observational side, Rupen (1991)
presents evidence that the non-thermal pressure gradients equal 400\% of
the thermal pressure gradient in NGC 891, and 50\% in NGC 4565.  Since
the luminous mass distributions in these galaxies are very similar, one
would expect very different gas layer widths if non-thermal effects were
indeed important.  In fact, these two galaxies have almost
indistinguishable flaring curves, and we can conclude that thermal
pressure gradients dominate \cite{mpR91,rpO96a}.

\begin{figure*}
   \epsfxsize 0.90\hsize
   \epsfysize 0.70\vsize
   \epsffile{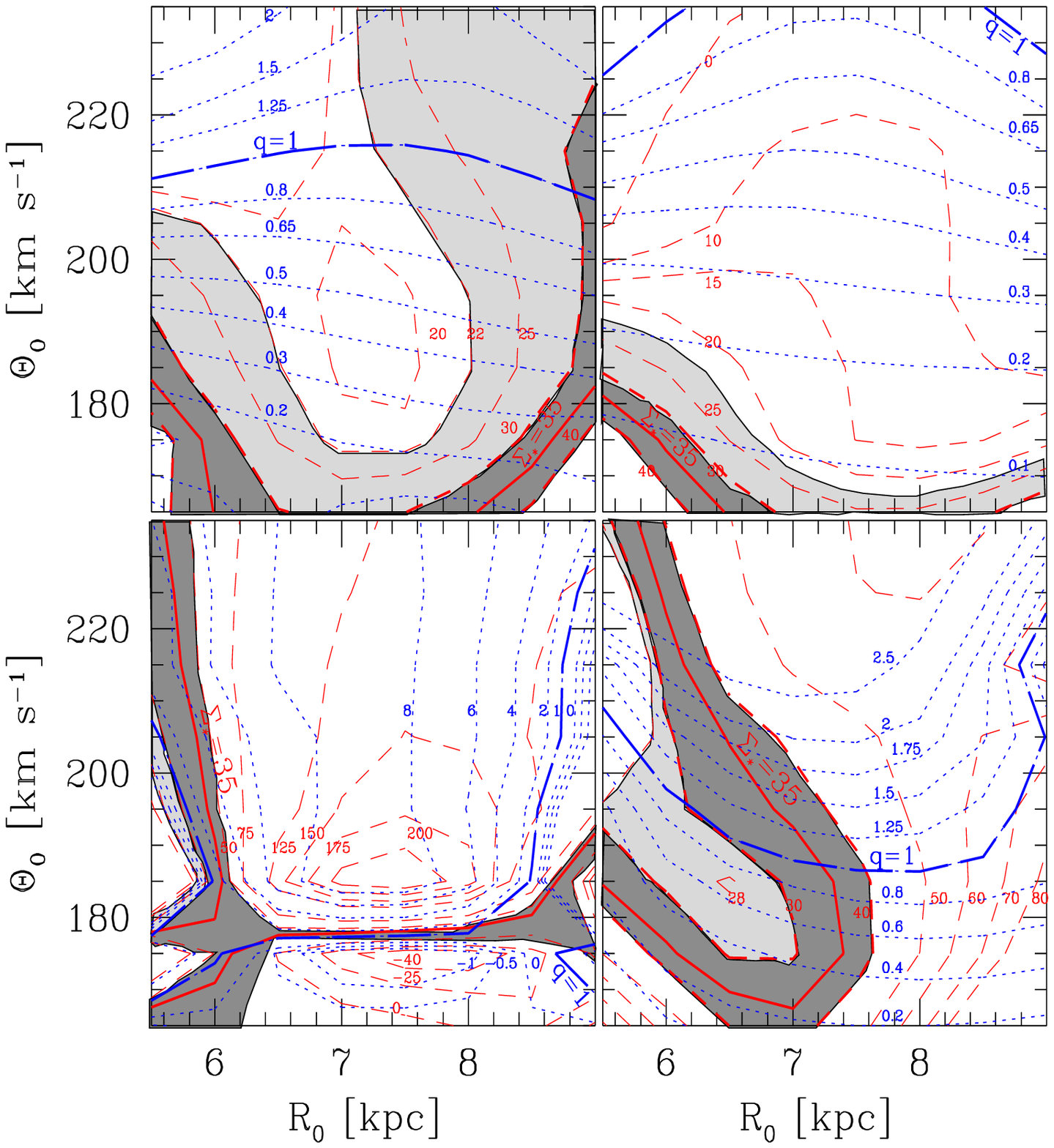} 

\caption{ \label{fig:Sstr_q_R0_T0_siggas} Here we present the {\em
mutually consistent} set of Galactic constants, stellar columndensity
in the Solar neighborhood (\protect\Sstr; long dashed lines) and halo
flattening ($q$; dotted lines).  The individual panels show the
results of the calculation for several values of
\protect\sgasp/\protect\sgas: 0.85 (\protect\sgasp=8 \protect\kms,
lower left), 1.0 (\protect\sgasp=9.2 \protect\kms, lower right), 1.15
(\protect\sgasp=10.5 \protect\kms, upper left), and 1.30
(\protect\sgasp=12 \protect\kms, upper right).  The oblate-prolate
boundary is indicated by the heavy dash-dotted line.  The heavy full
line and the heavy dashed line corresponds to KG89's determination of
\protect\Sstr, and the $\pm1-\sigma$ values.  The upper limit of
GBF97's determination of the stellar columndensity corresponds to the
\protect\Sstr $\sim30$ \protect\MSpcsq\ contour.  Both the halo
flattening and stellar column are determined to $\sim6$\% accuracy
(see also Appendix~\ref{app:Error_Estimation}).  Because the Galactic
constants, \protect\Sstr, and $q$ have to be mutually consistent one
cannot arbitrarily choose the four parameters.  Fixing one of the four
parameters severely restricts the other three.  Any two parameters
follow immediately from any choice of the other two, for a given
$\sigma_g$. In the shaded region of parameter space, the mass of the
stellar disk is as measured by KG89 (dark shading) and GBF97 (light
shading). }

\end{figure*}

To summarize, simplified theoretical considerations indicate that
non-thermal pressure support could dominate the hydrostatic balance if
extreme assumptions about the magnetic field geometry and the cosmic ray
energy density are made.  More realistic assumptions regarding the
vertical gradients of $P_B$ and $P_C$, as well as observational data
lead to the opposite conclusion.

In the next paragraph we will investigate the effects of significant
non-thermal pressure support by treating the gaseous velocity as an
unknown, notwithstanding the indications that non-thermal pressure
support is actually small in the outer Galaxy.  Based on the arguments
presented above, we expect \sgasp\ to lie between $\sim$0.85 and
$\sqrt{2}$ times the default value of 9.2 \kms, while a more realistic
upper limit to \sgasp/\sgas\ might be 1.05 [cf. 
eqn.~(\ref{eqn:sigrat_non_thermal_tR0_real})]. 

\mVskip{-2mm}
\subsection{Constraints on the pressure term}
 \label{sec:Constraints_on_the_pressure_term}

In this section we address the question as to how the interrelations
between the Galactic constants, \Sstr\ and $q$ are affected when
assuming that \sgas\ is unknown. We follow the procedure to determine
the halo flattening as outlined in the Introduction (and Paper~I) for
several trial values of \sgas\ ($\sgasp = 8.0, 8.6, 9.2, 9.9, 10.5,
12$ and 14 \kms), and on the same $R_0-\Theta_0$ grid as in
Paper~I. As predicted by equation~(\ref{eqn:qHI_T0_R0}), the inferred
halo flattening and stellar columndensity are rather sensitive to the
adopted value of \sgasp. In figure~\ref{fig:Sstr_q_R0_T0_siggas} we
present $q$ (dotted lines) and \Sstr\ (long dashed lines) as
calculated for increasing values of \sgasp, from left to right, and
from top to bottom.  The oblate region of parameter space, below the
heavy dash-dotted line, is strongly restricted if $\sgasp/\sgas < 1$
(lower-left panel), while essentially the whole range of the Galactic
constants is allowed if the effective dispersion is as large as 12
\kms\ (upper-right panel).  From these figures we can also infer that
the mass of the stellar disk, if measured accurately, constrains the
allowed range of \sgasp.

\begin{figure*}
   \begin{center}
   \epsfxsize 0.95\hsize 
   \epsfysize 0.85\vsize
   \epsffile{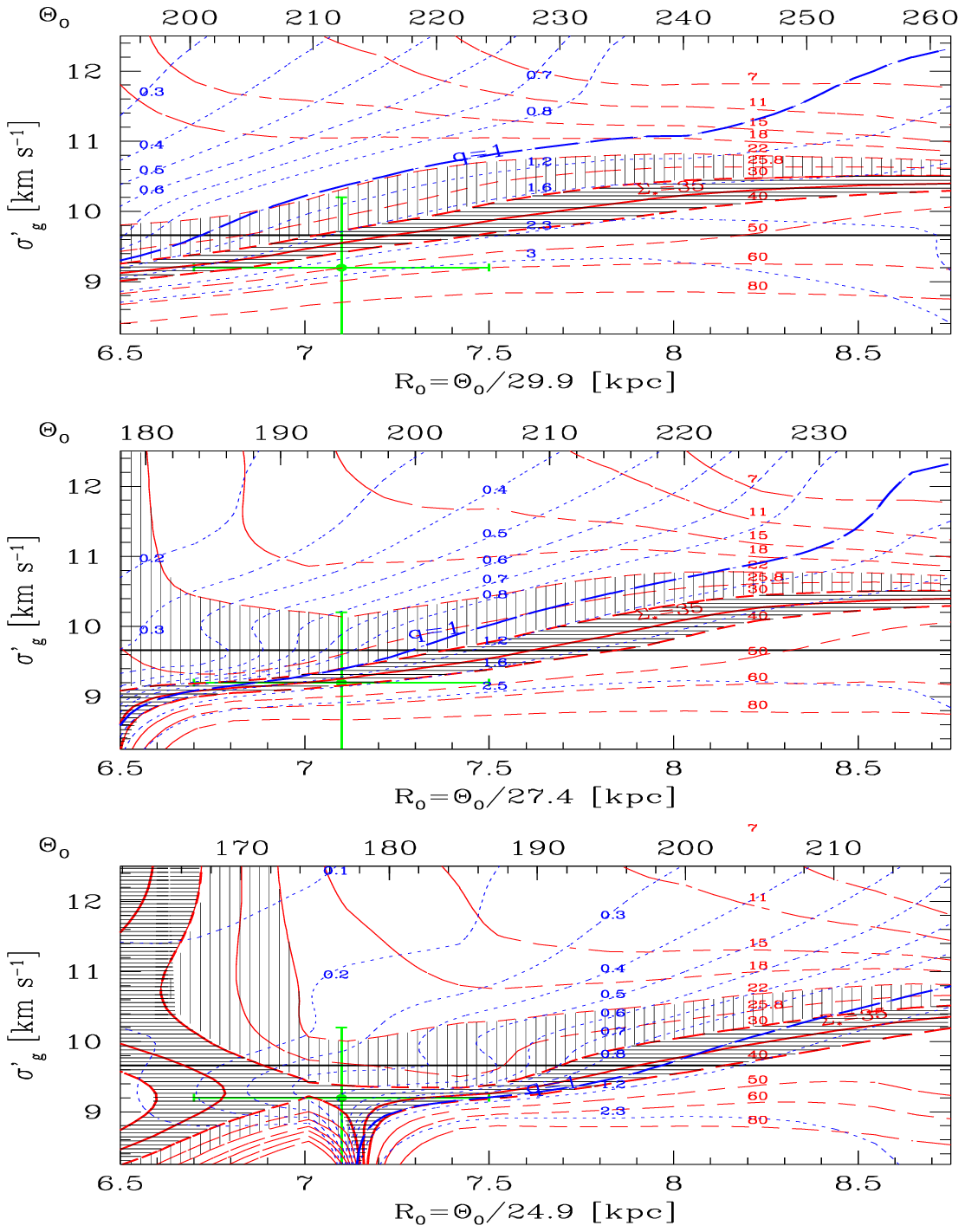} 

\caption{ \label{fig:Tq1_S_sgas} In this figure we present the
interrelations between the adopted value of the gaseous velocity
dispersion ($\sgas'$) and the Galactic constants.  The line coding is
the same as in Figs.~\ref{fig:Sstr_q_R0_T0_siggas}.  This figure was
generated by extracting the halo flattening and \protect\Sstr\ values
along the lines $\Theta_0/R_0 = 24.9, 27.4$, and 29.9 (from top to
bottom), for several values of the effective gaseous velocity
dispersion.  The fat cross represents our best estimate for the
Galactic constants [$R_0=7.1 \pm 0.4$ kpc, $\Theta_0=184 \pm 8$
\protect\kms, \protect\cite{rOmM98c}], and the gaseous velocity
dispersion [$\sgas = 9.2 \pm 1$ \protect\kms, \protect\cite{sM95}].
The thick horizontal line represents our theoretical expectation as to
the maximum value of \protect\sgasp\
(eqn.~\ref{eqn:sigrat_non_thermal_tR0_real}).  Large effective
velocity dispersions require flattened halos so that the model gas
layer widths are as thin as observed.  Round, and even prolate, halos
are found for small $\sgas'$ values.  In the shaded region of
parameter space, the mass of the stellar disk is as measured by KG89
(horizontal dark shading) and GBF97 (vertical light shading).  }

\end{center}
\end{figure*}

\nid Values for \sgasp\ that are as much as 30\% larger than the
default value (upper right panel of
Fig.~\ref{fig:Sstr_q_R0_T0_siggas}) are completely excluded because of
the very small stellar disk masses required.  On the other hand, lower
velocity dispersions need lower density (rounder) halos and hence a
more massive stellar disk, for given Galactic constants.  This is
indeed what is observed in the lower-left panel of
Fig.~\ref{fig:Sstr_q_R0_T0_siggas} for $\VSUN \ga 175$ \kms.  However,
below \VSUNn$\sim$175 \kms, the situation is much more chaotic
(extreme negative values for \Sstr\ and $q$), which is the result of
the fact that the \HI\ flaring and the stellar kinematical constraints
are mutually exclusive (Appendix~\ref{app:appendix_q_Sigma_Theta}).
In fact, an effective velocity dispersion $\la$ 8 \kms\ is essentially
ruled out.

\subsection{Other Constraints}
 \label{sec:Other_Constraints}

   In order to take all constraints properly into account, we should
present a three dimensional plot with \RSUNn, \VSUN and \sgasp\  as the
axes.  Because this is a little tedious, we opt for to
determine the halo flattening and \Sstr\  along lines of constant
$\Omega_0$.  We select $\Omega_0$ values which bracket
the uncertainty of the proper motion of SgrA$^*$ \cite{Rea99}.  In
each of the three panels of Figure~\ref{fig:Tq1_S_sgas} ($\Omega_0=29.9,
27.4, 24.9$ \kmskpc, from top to bottom) we present contours of
constant halo flattening by dotted lines as a function of \RSUN and
\sgasp. The heavy dashed-dotted line is the round-halo contour. The
hashed parts of the diagram depict regions of parameter space where
\Sstr = 35 \pmt 5 \MSpcsq\  (heavy horizontal hash) and \Sstr = 27.8
\pmt 3.8 \MSpcsq\  (light vertical hash). These columndensity ranges
correspond to the stellar disk mass as determined by KG89 and GBF97,
respectively.  In figure~\ref{fig:Tq1_S_sgas} we also plot our
estimate to the upper limit of the non-thermal pressure support
($\sgasp \sim1.05 \sgas$) as the thick horizontal line. The cross at
(\RSUNn,\sgasp) = (7.2,9.2) corresponds to the value of \RSUN derived
from the Oort constant \cite{rOmM98c} and the standard value of the
gaseous velocity dispersion.

A general feature of these diagrams is that flatter halos are found in
regions with large values of \sgasp.  This arises naturally from the
fact that a stronger gravitational pull is needed to constrain a gas
layer with additional pressure support, for a given observed thickness
[see also eqn.~(\ref{eqn:in_dark_halo})].  These figures also clearly
show that regions with large DM densities, due to either large \VSUN or
small $q$, have low stellar columndensity in the Solar neighbourhood. 
We can seen that significant non-thermal pressure support, which we
define here as having \sgasp = $\sqrt{2} \sgas \sim 13$ \kms, requires
very highly flattened dark matter halos {\em and} very small stellar
columndensities in the Solar neighbourhood.  Such strong non-thermal
pressure support is thus ruled out.

Several other generic conclusions can be drawn from
figure~\ref{fig:Tq1_S_sgas}. First, highly flattened halos are only
possible for realistic values of \Sstr\ if $R_0 \la 7$ kpc, and only
for small values of $\Omega_0$. Second, large values for \RSUNn, an
acceptable stellar columndensity, and an oblate halo occur only if the
angular velocity of the Milky Way is small, whatever the value of the
effective velocity dispersion. Third, the extra degree of freedom
associated with \sgasp\ does not greatly influence the inferred halo
flattening if a good constraint on \Sstr\ can be used. And finally,
the combined constraints set by the observed stellar columndensity
and the halo's oblateness severely restrict the allowed range for
\sgasp, in particular for $R_0 \ge 7$ kpc.

More specific results follow if we are willing to make more
restrictive assumptions.  For example, if we assume that $\RSUN \ge 7$
and $\sgasp \le 10.2$, we find $q \ga 1.0 \, (\Omega_0=29.9)$, $q \ga
0.65\, (\Omega_0=27.4)$, and $q \ga 0.2 \, (\Omega_0=24.9)$. In case
$\RSUN \ge 8$ kpc, $q \ga 2$, $q \ga 1.3$ and $q\ga 0.8$ for the same
angular velocities.  Alternatively, when choosing particular values
for the Galactic constants, simple relations between the remaining
three parameters of the mass model follow.  For example, taking the
IAU-recommended values for \RSUN\ and \VSUN\ (8.5 kpc, and 220 \kms)
we find $q \sim 2.26 - 0.90 \times d\sgasp + 0.11 \times d\sgasp^2$
{\em and} $\Sstr \sim 65.6 -35.2 \times d\sgasp + 4.8 \times d\sgasp^2$, 
with d\sgasp=\sgasp-9.2. Also, with the standard value for the
gaseous velocity dispersion we find: $\Theta_0(q\le1) \la 187 + 5
\times dR^2_0$ {\em and} $\Sstr(0.5 \la q \la 1) \sim 
37.5 + 18.4\times dR_0 + 8.5\times dR^2_0$, with d\RSUN=\RSUNn-7.5
(cf.  Fig.~\ref{fig:Sstr_q_R0_T0_siggas}, the lower-right panel).

An inspection of the lower two panels of figure~\ref{fig:Tq1_S_sgas}
reveals that rather tight constraints can be placed on the halo shape,
the local angular velocity and the effective velocity dispersion if
the halo is oblate and $30 \la \Sstr \la 40$ \MSpcsq\ and $R_0 \ge 7$
kpc. In that case we find: $24.9 \la \Omega_0 \la 27.4$ \kmskpc,
$0.5\la q \le 1$, and $8.6 \la \sgasp \la 10.3$ \kms. If we impose the
additional constraint that the non-thermal pressure support is limited
to 5\% as derived in equation~(\ref{eqn:sigrat_non_thermal_tR0_real}),
it follows that the Sun's distance to the Galactic centre is less than
8 kpc, and that the rotation speed of the Milky Way is less than 200
\kms\ at the Solar circle.

The stellar disk mass provides a strong constraint on the effective
velocity dispersion of the gas: a low disk mass requires more dark
matter, which would lead to a thinner gas layer in the outer Galaxy if
\sgasp\ were not increased. For example, if the stellar disk mass
exceeds 22 \MSpcsq, figure~\ref{fig:Tq1_S_sgas} shows that the
effective velocity dispersion has an upper bound of about 10.5 \kms,
so that the non-thermal pressure support can not exceed 14\%.

\mVskip{-4mm}
  \section{Summary and Conclusions}
\label{sec:Summary_and_Conclusions}

In Paper~I we showed that the constraints on the dark matter density
in the Solar neighbourhood and the outer Galaxy place tight limits on
the choice of the Galactic constants, the mass of the stellar disk,
and the dark halo's flattening.  The internal errors for this
procedure are of order 6\% for both the halo flattening and the local
stellar column density (see Appendix \ref{app:Error_Estimation}).

In this paper we present the analytical tools that allow for an
efficient search through parameter space.  Employing these tools, we
extend the analysis of Paper~I by investigating the effects of the
ill-determined contribution of non-thermal pressure support on the
\HI\ flaring, and the consequences for the inferred \RSUNn, \VSUNn,
\Sstr, and $q$ values.  The strongest constraints available are the
observed \VSUNn/\RSUN ratio, the mass of the stellar disk and the fact
that the dark halo is almost certainly oblate.  Taken together, these
constraints rule out substantial contributions to the support of the
\HI\ layer by cosmic ray pressure or magnetic fields if the Sun's
distance to the Galactic centre is greater or equal than 7 kpc, in
agreement with theoretical predictions.  We find that the Milky Way's
dark matter halo is close to spherical for all but the smallest values
of \RSUN or \VSUNn.  A dark matter halo as flattened as $q=0.2$ is
only possible if our distance to the Galactic centre is smaller than
about 6.8 kpc.

It is possible to construct a self-consistent oblate model of
the Galaxy with \RSUN = 8.5 kpc and \VSUN = 220 \kms, but only if the
local stellar columndensity is less than about 18 \MSpcsq, {\em and}
$\sgasp \ga 11$ \kms.

Kuijken \& Gilmore's (1991) determination of the columndensity of
matter within 1.1 kpc of the plane (71 \pmt 6 \MSpcsq) is robust and
valid over a wide range of Galactic constants and disk scale-lengths.
For the stellar contribution to this total mass we suggest a {\em
consensus} average of $\Sstr = 35$ and a {\em consensus} error of 10
\MSpcsq.

If \RSUN $\ga$ 7 kpc, then the dark halo of the Milky Way is fairly
close to spherical, independent of the amount of non-thermal pressure
support.  Such a round halo argues against dissipational baryons as a
viable dark matter candidate.  Further, since all other baryonic dark
matter candidates have already been observationally excluded
\cite{HO86}, we must conclude that the dark matter halo of the Milky
Way is most likely made up of something altogether more exotic
(MACHOs, neutrinos, axions, neutralinos ...).  Even in the Solar
neighborhood, there are non-negligible quantities of this material:
our proposed dark halo models imply that it amounts to some 0.42
GeV/c\rtp{2} per cubic centimetre or (11 \pmt 5) m\MSpccub.  The
direct detection of this material remains a challenge for experimental
physicists.

Employing the local disk mass and the flaring of the Galactic \HI\
layer, we find strong correlations between the parameters of
axisymmetric mass models.  We presented several examples in the
previous section.  Since we make strong predictions as to the values
of \RSUNn, \VSUNn, \Sstr, $q$ and \sgasp\
(Figs.~\ref{fig:Sstr_q_R0_T0_siggas} and \ref{fig:Tq1_S_sgas}), our
models can be subjected to experimental verification.  For example,
all parameters but \sgasp\ could be determined using astrometric data
from future astrometric space missions such as FAME, SIM and GAIA.
Such high precision data are ideally suited to support, or falsify,
the models we propose here.  It would also be worthwhile to
investigate the effects of deviations from axisymmetry on the inferred
halo shape and rotation speed. Whatever the outcome, we will learn a
great deal more about the structure and dynamics of the Milky Way
galaxy.

\mVskip{-5mm}
\section*{acknowledgments}

We thank Andy Newsam, Irini Sakelliou, Konrad Kuijken, Marc
Kamionkowski, Jacqueline van Gorkom, Chris McKee, and James Binney for
useful discussions.  We also wish to thank the referee for valuable
suggestions for improvements.  Much of this research was performed by
the authors while at the University of Southampton and while RPO was
at Rutgers University.  RPO thanks Columbia University's Astronomy
department for its hospitality.  This research has made use of NASA's
Astrophysics Data System Abstract Service.

\appendix

\mVskip{-5mm}
\section{More on the vertical disk-halo conspiracy}
\label{app:appendix_Kz}

In section~\ref{sec:The_local_stellar_column_density_a_constraint} we
reviewed the stellar kinematical route to determine the local stellar
columndensity.  In this appendix we try to emulate and check this method
in some detail.  In order to do so we calculate the exact vertical force
law $K_{\rm z,exact}$ for a limited number of Galaxy models listed in
Table~\ref{tab:Disk_Halo_parameters} using the global approach
(\S~\ref{sec:Constraints_from_the_thickness_of_the_gas_layer}; cf. 
Olling 1995).  We also compute an approximation to $K_{\rm z,exact}$
from the local mass distribution
(\S~\ref{sec:Constraints_from_the_thickness_of_the_gas_layer}; cf. 
Olling 1995), which follows from an integration of the Poisson equation:

\negSMLskip
\begin{eqnarray}
K_{\rm z,local}(z)
   &=& -4 \pi G \int_0^z  dz'  \left[ \rho_m(z') + \rho_{rot} \right] \, ,
   \label{eqn:eqn_for_Kz_local}
\end{eqnarray}

\noindent with $\rho_m$ the matter density, $\rho_{rot} (\equiv
\frac{-2}{4 \pi G} \frac{V_{rot}}{R} \frac{dV_{rot}}{dR})$ a pseudo
density which arises from gradients in the rotation curve, and \Kz{}
the vertical force per unit mass.

\begin{figure*}
   \epsffile{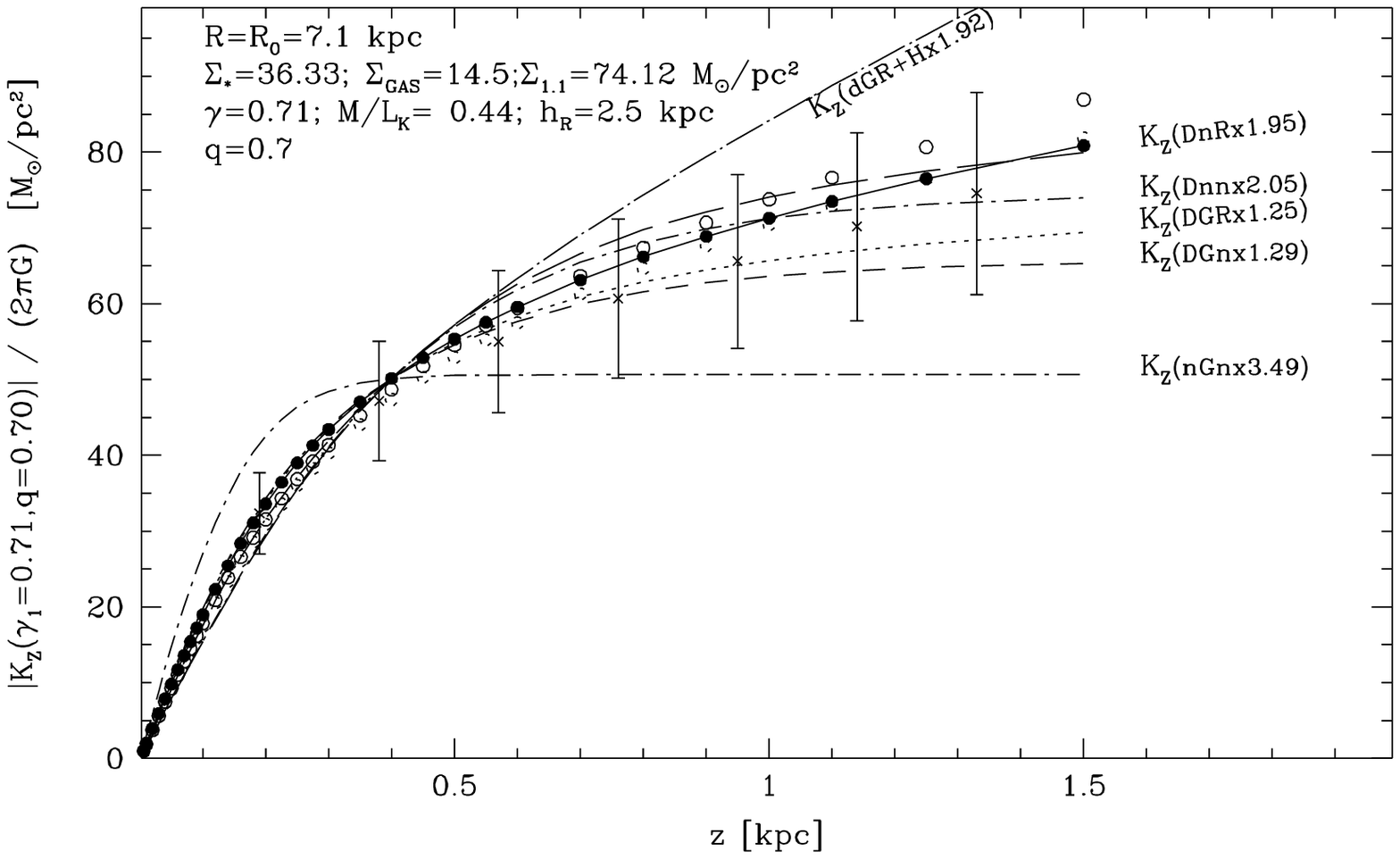}

\caption{ \label{fig:Kz_laws} We present the exact vertical force
(filled circles) at Galactocentric distance \protect\RSUN for a model
that almost reproduces the values for \protect\Sstr\ and \protect\SKG\
proposed by Kuijken \& Gilmore (1989, 1991).  The local approximation
to the exact force law that uses the slope of the rotation curve (open
circles) or assumes a flat rotation curve (dashed open circles) are
also indicated.  Some other models in which the DM is distributed like
various superpositions of the known components are also plotted: 1)
$\rho_{DM} = 1.29 \times (\rho_*+\rho_g)$ (labeled ``DGnx1.29'); 2)
$\rho_{DM} = 1.25 \times (\rho_*+\rho_g+\rho_{rot}$) (``DGRx1.25'');
3) $\rho_{DM} = 2.05 \times \rho_*$ (``Dnnx2.05''); 4) $\rho_{DM} =
1.95 \times (\rho_*+\rho_{rot}$) (``DnRx1.95''); 5) $\rho_{DM} = 3.49
\times \rho_g$ (``nGnx3.49'').  For the curve labeled ``dGR+Hx1.92'' we
used a smaller columndensity of stars 25.8 \protect\MSpcsq\
\protect\cite{GBF97}, and multiplied the halo contribution by 1.92. 
Most of the models have very similar force laws in the region close to
the plane (see also Table~\protect\ref{tab:Dark_matter_types}).  We also
include KG89's vertical force law and the errors thereupon (crosses with
error bars). 
}

\end{figure*}

In figure~\ref{fig:Kz_laws} we present true vertical force (filled
circles) and the local approximation (open circles) for one particular
combination of parameters which lie close to the best estimates for
the Milky Way \cite{kKgG8991,rOmM98c}.  We see that in the Solar
neighborhood and close to the plane, the local approximation is
reasonably accurate\footnote{For the ensemble of models for which the
parameters differ by $\le$ 1-$\sigma$ from the model presented here,
the approximations based on equation (\ref{eqn:eqn_for_Kz_local})
reproduce the true \Kz{} typically to within {\bf 7\%}: sometimes
models that include the rotation curve gradient term perform best,
sometimes not. The same is true for the models listed in
Table~\ref{tab:Disk_Halo_parameters} with other Galactic constants
than those presented in figure~\ref{fig:Kz_laws}.}  and can be used as
an approximation to the true vertical force law.  Significant
deviations from the true \Kz\ only occur above approximately 800
pc. We also include the observationally determined \Kz{} [crosses with
error bars; KG89's equation (1) with D=250 pc].

Now we investigate the possibility that the actual dark matter density
law differs from the distribution in our models [eqn. 
(\ref{eqn:rho_halo_Rz})].  Choosing a different DM distribution while
keeping the observed \Kz{} fixed requires a different luminous mass
distribution.  As a simple test case, we plot a model where the DM
density is multiplied by 1.92 (labeled ``dGR+Hx1.92'' in
Fig.~\ref{fig:Kz_laws}).  In this case, we have to decrease the stellar
columndensity to the GBF97 value (25.8 \MSpcsq) in order to keep \Kz{}
approximately unchanged.  Next we consider DM distributions inspired by
Bahcall's ``P'' models (1984b), where the DM density is assumed to be
a linear combination of the known components:

\begin{eqnarray}
\rho_{err} \hspace*{-2mm} &=& \hspace*{-2mm}
   X_d    \times \rho_d + 
   X_g    \times \rho_g + 
   X_h    \times \rho_h +
   X_{rot}\times\rho_{rot}  \, ,
   \label{eqn:P_models}
\end{eqnarray}

\noindent where the subscripts $d, g, h$ denote the disk, gas, and
halo components, respectively.  Using this formalism, we can calculate
the columndensities of all components (ISM, stars, DM and \SKG) which
arise from a particular choice of the $X_i$-values.  From such an
erroneous density distributions ($\rho_{err}$) we determine the
erroneous vertical force ($K_{z,err}$) using
equation~(\ref{eqn:eqn_for_Kz_local}).  We present the
$K_{z,true}/K_{z,err}$ ratio, averaged in several z-height ranges, in
Table~\ref{tab:Dark_matter_types} (columns 5-7).  In this table we
also list the $X_i$, \Sstr\ and \SKG\ values for some models.  In
addition to the P-models, we also evaluate models in which we keep the
ISM and $\rho_{rot}$ contributions fixed while varying the stellar and
dark matter densities such that the force at 400 pc above the plane
equals the true value.  From table~\ref{tab:Dark_matter_types} we see
that it is possible to decrease \Sstr\ by more than an order of
magnitude and simultaneously increase $X_h$ without altering $K_z(|z|
\le 800$pc) substantially.  It seems that the vertical distribution of
dark and stellar matter conspire in such a way that their relative
contributions can not be easily separated.  Thus, this ``vertical
disk-halo conspiracy'' is much like the classical disk-halo conspiracy
which arises in considerations of the radial distribution of luminous
and dark luminous matter derived from galaxy rotation curves.

\begin{table}
 \caption{The accuracy to which various erroneous density distributions
can reproduce the true vertical force.  The first four columns describe
which $\rho(z)$ was used [cf.  eqn.~(\protect\ref{eqn:P_models})].  The
next three columns indicate the average ratio of $K_{\rm z,true}/K_{\rm
z,err}$ over the vertical ranges indicated [0 - 0.4 kpc (Z1), 0.4 - 0.8
kpc (Z2), and 0.8 - 1.2 kpc (Z3)].  The typical variation within these
radial ranges is a few \%, except for the gas-only case where it can be
up to 35\%.  We also indicate the model values for \protect\Sstr\  and
\protect\SKG, in units of \protect\MSpcsq.  The last two lines give the
average and standard deviations of the lines above (excluding the
gas-only model). 
}
 \label{tab:Dark_matter_types}
\begin{tabular}{|c|c|c|c|c|c|c|c|c|}
$X_d$&$X_g$ &$X_{rot}$ &$X_h$ & Z1 & Z2  & Z3    &\Sstr & \SKG  \\ \hline
1.25 & 1.25 & 1.25 & 0.00 & 1.03 & 0.99  & 0.93  & 45.4 &  63.6 \\
1.29 & 1.29 & 0.00 & 0.00 & 0.95 & 1.04  & 1.17  & 46.9 &  65.6 \\
2.05 & 0.00 & 0.00 & 0.00 & 0.85 & 1.03  & 1.01  & 74.5 &  74.5 \\
1.95 & 0.00 & 1.95 & 0.00 & 0.85 & 1.03  & 1.04  & 70.8 &  70.8 \\
0.00 & 3.49 & 0.00 & 0.00 & 1.35 & 0.90  & 0.72  &  0.0 &  50.6 \\ \hline
1.40 & 1.00 & 1.00 & 0.03 & 1.03 & 0.99  & 0.93  & 50.9 &  66.1 \\
1.25 & 1.00 & 1.00 & 0.46 & 1.03 & 0.99  & 0.93  & 45.4 &  70.6 \\
1.00 & 1.00 & 1.00 & 1.17 & 0.97 & 1.02  & 1.08  & 36.3 &  78.1 \\
0.74 & 1.00 & 1.00 & 1.92 & 0.95 & 1.04  & 1.17  & 26.9 &  86.1 \\
0.50 & 1.00 & 1.00 & 2.59 & 0.93 & 1.06  & 1.24  & 18.2 &  92.9 \\
0.25 & 1.00 & 1.00 & 3.30 & 0.92 & 1.08  & 1.33  &  9.1 & 100.4 \\ 
0.12 & 1.00 & 1.00 & 3.67 & 0.91 & 1.09  & 1.37  &  4.4 & 104.3 \\ \hline
     &      &      &      &      &       &       &39\pmt&  77\pmt    \\ 
     &      &      &      &      &       &       &58\%  &  21\%\\ \hline
\end{tabular}
\end{table}

From the data presented in Table~\ref{tab:Dark_matter_types} it is
clear that, without additional assumptions, stellar kinematical data
extending to a few hundred parsec are not sufficient to determine the
mass of the stellar disk.  However, \SKG\ is much better determined
(rightmost column of Table~\ref{tab:Dark_matter_types}), in agreement
with Kuijken
\& Gilmore (1991).  In order to discriminate between models, it is
essential to incorporate a rotation curve constraint to limit the
possible DM densities.  Kuijken \& Gilmore (1989) used such a constraint
explicitly, while Bahcall (1984b) and Flynn \& Fuchs (1994) used
``reasonable'' values for the density of non-disk-like dark matter to
rule out extreme values for the mass of the stellar disk.  Furthermore,
if data at larger $z$ are considered, the differences between the force
laws shown in figure \ref{fig:Kz_laws} start to become apparent and can
be used to limit the allowed range for \Sstr\ (Flynn \& Fuchs 1994). 

Although the the model parameters we have used to construct
Figure~\ref{fig:Kz_laws} and Table~\ref{tab:Dark_matter_types} lie at
the extreme end of the models investigated by Kuijken \& Gilmore, we
find that other models, within the 1-2$\sigma$ range of \VSUNn/\RSUN,
\hd, \Sstr, and \SKG, yield similar results.  Thus, mass models for the
Milky Way {\em have} to conform to the constraint \SKG = (71 \pmt 6)
\MSpcsq.  To summarize the above, in order to determine the stellar
columndensity from stellar kinematics, the existence of the vertical
disk-halo conspiracy requires that one {\em has} to include
self-consistent rotation curve constraints and/or sample the region
above 800 pc.

\mVskip{-5mm}
\section{Stellar kinematics and the relation between the model's parameters}
  \label{app:appendix_q_Sigma_Theta}

In section \ref{sec:The_connection_between_dark_and_luminous_matter} we
described how the observationally determined \SKG-value imposes
correlations between luminous and dark matter in the Solar
neighbourhood.  In this section we present some specific examples to
illustrate what we can learn about the structure of the Milky Way by
treating \Sstr\ as a parameter which is only constrained by \SKG\ [cf. 
eqn~(\ref{eqn:SKG_his})].  Note that these correlations are independent
of any \HI\ flaring constraints. 

At the Solar circle, and for $z = 1.1$ kpc, equation
(\ref{eqn:Sigma_DM1p1_Vhalo}) can be rearranged to read:

\negSMLskip
\begin{eqnarray}
\frac{\SHkg(q)}{\SHkg(1)} 
   \hspace*{-1em} &\approx&  \hspace*{-1em}
   1.1 \RSUN + \RSUN (0.9-\RSUNn) \, q 
   \hspace*{8mm} q\la 0.15 \label{eqn:Sq_la} \\
   &\approx&  q\rtp{-13/20} \, ,
   \hspace*{3.1cm} q\ga 0.15 \label{eqn:Sq_ga}
\end{eqnarray}

\noindent where the core radius has been set to \RSUNn.  In this
appendix we use --for the sake of convenience-- the symbol $q$ for the
halo flattening derived from the local stellar kinematics constraint,
and \qHI\ for the halo's shape derived from the \HI\ flaring. Taking
the fiducial values for \SKG, \Sstr, and \Sgas\ to be 71, 35, and 14.5
\MSpcsq, respectively, and employing equation~(\ref{eqn:SKG_his}) we
determine the fiducial value for \SHkg\ to be 21.5 \MSpcsq.  Further,
defining $\Sigma_* = 35 + \delta\Sigma_*$, use the fiducial
columndensities defined above, and apply equations (\ref{eqn:Sq_la})
and (\ref{eqn:Sq_ga}), we can relate the required halo flattening to
$\delta\Sigma_*$:

\negSMLskip
\begin{eqnarray}
q \hspace*{-2mm} &\approx&  \hspace*{-2mm}
   \frac{21.5 - \RSUN \SHkg(1) - \delta\Sigma_*}
        {0.9R_0^2 \SHkg(1)}
  \label{eqn:q_la} \\
  \hspace*{-2mm} &\approx& \hspace*{-2mm}
  \left( \frac{\SHkg(1)}{21.5} \right)^{20/13}
  \left[
     1 + \left(\frac{ 20/13 }{21.5}\right) \delta\Sigma_*
  \right]
  \label{eqn:q_ga}
\end{eqnarray}

\noindent for the same $q$-ranges as in Eqns.  (\ref{eqn:Sq_la}) and
(\ref{eqn:Sq_ga}).  Since $\SHkg(q=1)$ depends on \VSUN [eqn. 
(\ref{eqn:Sigma_DM1p1_Vhalo})], the solutions of equations
(\ref{eqn:q_la}) and (\ref{eqn:q_ga}) depend on \RSUNn, \VSUNn, and
$\delta\Sigma_*$.  We use our mass models to determine $\SHkg(q=1;
R_0,\Theta_0)$ and rewrite equations (\ref{eqn:q_la}) and
(\ref{eqn:q_ga}) for a few interesting cases.  For $\delta\Sigma_*=0$ we
find:

\negSMLskip
\begin{eqnarray}
q \hspace*{-2mm} &\approx& \hspace*{-2mm}
   0.07   \pm 0.05                   \hspace*{2.5cm} (\VSUNn=165)
   \label{eqn:q_V165} \\
q \hspace*{-2mm} &\approx& \hspace*{-2mm}
   0.4 - 0.1 (\RSUN-7.1)   \pm 0.1 \hspace*{0.5cm} (\VSUNn=175)
   \label{eqn:q_V175} \\
q \hspace*{-2mm} &\approx& \hspace*{-2mm}
   0.8 - 0.2 (\RSUN-7.1)   \pm 0.2 \hspace*{0.5cm} (\VSUNn=185)
   \label{eqn:q_V185}
\end{eqnarray}

\noindent Thus, lower rotation speeds and larger \RSUNn's require more
highly flattened halos.  Furthermore, for any fixed value of \RSUNn, the
{\em slope} of the $q-\delta\Sigma_*$ relation depends strongly upon
\VSUNn.  For example, with \RSUNn=7.1 kpc we find:

\negSMLskip
\begin{eqnarray}
q \hspace*{-2mm} &\approx& \hspace*{-2mm}
   0.069 + 0.005 \, \delta\Sigma_*  \hspace{1cm} (\VSUNn=165)
   \label{eqn:R07p1_T0165} \\
q \hspace*{-2mm} &\approx& \hspace*{-2mm}
   0.400 + 0.030 \, \delta\Sigma_*  \hspace{1cm} (\VSUNn=175)
   \label{eqn:R07p1_T0175} \\
q \hspace*{-2mm} &\approx& \hspace*{-2mm}
   0.800 + 0.060 \, \delta\Sigma_*  \hspace{1cm} (\VSUNn=185)
   \label{eqn:R07p1_T0185}
\end{eqnarray}

\noindent Equations (\ref{eqn:q_V165}) through (\ref{eqn:R07p1_T0185})
clearly show that small values of the Galactic rotation speed imply a
highly flattened dark matter halo, whatever our distance to the Galactic
centre, and whatever the mass of the stellar disk.  We also see that the
last relations constrain \Sstr\ rather tightly: for models with
$\VSUN \sim$ 185 (175) \kms, an increase in the stellar columndensity of
$\sim$ 3 (20) \MSpcsq\ covers the whole allowed range for $q$.

In section~\ref{sec:Constraints_on_the_pressure_term} and
figure~\ref{fig:Sstr_q_R0_T0_siggas} we have see the stellar kinematics
and \HI\ flaring constraints provide mutually exclusive constraints on
\Sstr\ and $q$ if both the rotation speed and the effective velocity
dispersion are small.  This can be understood as follows: equations~
(\ref{eqn:q_V165})-(\ref{eqn:R07p1_T0185}) show that the \SKG\
constraint implies small $q$-values for low \VSUNn's.  Likewise, the
\HI\ flaring constraint yields small $q$'s for small \VSUNn's, but only
if the velocity dispersion does not decrease by too much [cf. 
eqn.~(\ref{eqn:q_zeta_Theta_sigmaG})].  Furthermore, because the slope
of the $q(\Sstr)$ relation becomes shallower with decreasing \VSUNn, the
accessible range for \qdT\ decreases with \VSUNn.  If the flaring
analysis leads to a $\qHI\sim1$ halo, extreme values for \Sstr\ are
required to match \qdT\ with \qHI.  Obviously, if the required stellar
column exceeds \SKG, it is not possible to construct a self-consistent
model for that particular combination of Galactic constants and gaseous
velocity dispersion.  In these circumstances it is possible that our
procedure to determine \Sstr\ and $q$ yields extreme values for \Sstr\
and $q$, and sometimes even negative values. 

\mVskip{-5mm}
  \section{Error Estimation}
\label{app:Error_Estimation}
At this point it is also possible to estimate the accuracy to which
the various parameters in figure~\ref{fig:Sstr_q_R0_T0_siggas} are
determined. For example, as we have seen before, the halo flattening
inferred from the \HI\ flaring has only a slight dependency on the
mass of the stellar disk (cf. figure~3 of Paper~I). This allows us to
estimate the flattening of the halo by averaging the \qHI\ values from
the various model runs at a given \RSUN and \VSUNn. For example, the
nine \qHI\ errors tabulated in table~\ref{tab:Disk_Halo_parameters}
yield $\delta \qHI/\qHI \sim 0.06$ for the two low \RSUN values at
$q\sim 0.7$ Employing $q\propto \Theta^2_0$
[cf. eqn.~(\ref{eqn:qHI_T0_R0})] we find:
\begin{eqnarray}
   \frac{\delta \Theta_0}{\Theta_0} \sim \frac{\delta \qHI}{2\qHI} \, .
   \label{eqn:qHI_T0_errors}
\end{eqnarray}
The \qHI\ values as determined from the flaring measurements are thus
precise to about 6\%, while the value of the local Galactic rotation
speed we derive from \qHI\ measurement has an estimated accuracy of 3
percent.

To estimate how well the local stellar column density is determined
from the flaring and the observed \SKG\ value of the total column
density we re-write equation~(\ref{eqn:q_ga}) to read $q = c_1
(1+c_2\delta \Sigma_*)$, and find:
\begin{eqnarray}
\delta (\delta \Sstr) &\sim& \frac{q}{c_1 c_2}
   \sqrt{ (\frac{\delta q}{q})^2 + (\frac{\delta c_1}{c_1})^2 } \, ,
   \label{eqn:Sigma_S_errors}
\end{eqnarray}
where $c_1=\SHkg(1)/21.5$, $c_2=(20/13)/21.5$ and $q/(c_1 c_2)\sim
11.2$ for $q=0.8$ and \SHkg(1)=21.5.  If assume that $c_1$ is without
error, the second term in eqn.~\ref{eqn:Sigma_S_errors} vanishes, and
we arrive at a lower limit for the error in \Sstr\ of about 0.7
\MSpcsq. If we assign the full error in the observed total column
density (6 \MSpcsq) to \SHkg(1), we have $c_1 \sim 1 \pm 0.43$, so that
$\delta (\delta \Sstr)) = 11.2
\sqrt{(0.06)^2 + (0.43)^2} \sim 4.8 \MSpcsq$. We thus estimate that our
method of deriving the stellar column density has an accuracy somewhere
between 0.7 and 4.8 \MSpcsq\ or about to 2 to 14 percent.

The error estimates above are close to the errors derived from our
detailed modeling procedure.

\mVskip{-5mm}
\section{The ISM of the Milky Way Tabulated}
  \label{app:ISM}
In this appendix we present a tabulated version of the radial variation
of the atomic and molecular hydrogen, as well as their widths.  We
present these data in a manner which is {\em independent} of the values
of the Galactic constants as well as the shape of the rotation curve. 
The \HI\ columndensity at the Solar position includes 1.4 \MSpcsq\ of
ionized hydrogen.  No ionized hydrogen is included at any other radius. 
The columndensities listed do not include the contribution due to
Helium.  In our model calculations of the potential we increase the
listed columndensities to include 23.8\% Helium.  We have brought the
data from the sources listed in section \ref{sec:The_Mass_Components}
onto a common distance scale defined by Merrifield's (1992)
determination of the $W(R/R_0)=v_{rad}/(\sin{\ell}\cos{b})$ curve.  Here
$v_{rad}$ and $\ell$ and $b$ are the radial velocity and the Galactic
coordinates, respectively.  In practice this re-scaling works as
follows: 1) from the rotation curve [$\Theta'(R')$] in the original
reference, determine $R'/R'_0$ and $W'(R'/R'_0) \equiv R'_0
[\Theta'(R')/R' -\Theta'_0/R'_0]$ [the primed quantities refer to the
values assumed in the reference], 2) for each property $X'$ (gas layer
width and volume density), determine $X'(W')$ from $X'(R')$ and
$W'(R')$, 3) find Merrifield's $R/R_0$ values for which $W=W'$, 4) the
property $X'$ re-gridded onto Merrifield's distance scale is now given
by $X(R/R_0)$.

\begin{table}
\caption{The surface density and thickness of the atomic and
 molecular hydrogen components in the Milky Way as a function of
 scaled Galactocentric radius, $R/\RSUN$. The \HI\ and \Ht\
 columndensities have units \protect\MSpcsq. The widths listed are
 full widths at half maximum (FWHM) and have units of $R_0$. For
 $R>R_0$, the widths of the \protect\HI\ layer is the average of
 several works, see Paper~I for details. Some entries have been left
 blank: we were unable to derive reliable widths at these radii. The
 columndensities at these radii are from Wouterloot \etal (1990).}
\label{tab:ISM}
\begin{tabular}{|c|c|c|c|c|c}
 & & & & &  \\ \hline
$R/\RSUN$ & 
$\Sigma_{\HI}$ & 
$\Sigma_{\Ht}$ & 
${W}_{\HI}$ &
${\delta W}_{\HI}$ &
${W}_{\Ht}$ \\ \hline \hline
 0.218 & 0.59 & 1.08 & 0.0330 & 0.0014 & 0.0151 \\
 0.276 & 1.04 & 1.54 & 0.0320 & 0.0031 & 0.0158 \\
 0.309 & 1.23 & 1.47 & 0.0323 & 0.0017 & 0.0134 \\
 0.343 & 1.30 & 1.67 & 0.0306 & 0.0014 & 0.0119 \\
 0.376 & 1.29 & 2.67 & 0.0294 & 0.0008 & 0.0131 \\
 0.409 & 1.31 & 4.29 & 0.0268 & 0.0045 & 0.0140 \\
 0.440 & 1.41 & 4.91 & 0.0290 & 0.0031 & 0.0142 \\
 0.467 & 1.59 & 5.02 & 0.0313 & 0.0028 & 0.0143 \\
 0.499 & 1.85 & 5.57 & 0.0278 & 0.0011 & 0.0153 \\
 0.530 & 2.10 & 6.23 & 0.0240 & 0.0017 & 0.0165 \\
 0.559 & 2.30 & 5.87 & 0.0301 & 0.0014 & 0.0168 \\
 0.586 & 2.48 & 5.48 & 0.0346 & 0.0014 & 0.0163 \\
 0.612 & 2.90 & 5.18 & 0.0344 & 0.0023 & 0.0152 \\
 0.641 & 3.72 & 4.62 & 0.0415 & 0.0034 & 0.0139 \\
 0.668 & 4.36 & 3.94 & 0.0504 & 0.0017 & 0.0129 \\
 0.692 & 4.24 & 3.73 & 0.0525 & 0.0014 & 0.0120 \\
 0.717 & 3.60 & 3.97 & 0.0447 & 0.0031 & 0.0115 \\
 0.742 & 3.17 & 4.27 & 0.0452 & 0.0039 & 0.0122 \\
 0.764 & 3.23 & 4.33 & 0.0485 & 0.0130 & 0.0136 \\
 0.786 & 3.68 & 4.09 & 0.0495 & 0.0101 & 0.0152 \\
 0.806 & 4.22 & 3.60 & 0.0565 & 0.0039 & 0.0161 \\
 0.828 & 4.72 & 2.95 & 0.0530 & 0.0048 & 0.0159 \\
 0.848 & 4.90 & 2.44 & 0.0570 & 0.0034 & 0.0144 \\
 0.865 & 4.80 & 2.06 & 0.0572 & 0.0011 & 0.0134 \\
 0.882 & 4.52 & 1.70 & 0.0509 & 0.0014 & 0.0136 \\
 0.897 & 4.41 & 1.45 & 0.0525 & 0.0020 & 0.0146 \\
 0.913 & 4.75 & 1.30 & 0.0523 & 0.0036 & 0.0158 \\
 0.926 & 5.39 & 1.26 & 0.0652 & 0.0017 & 0.0167 \\
 0.938 & 6.13 & 1.29 & 0.0730 & 0.0020 & 0.0174 \\
 1.000 & 9.25 & 1.80 & 0.0577 & 0.0042 & 0.0199 \\
 1.058 & 8.57 & 1.80 & \ptn   & \ptn   & \ptn   \\
 1.089 & 7.77 & 1.28 & 0.0611 & 0.0044 & 0.0238 \\
 1.193 & 6.09 & 0.44 & \ptn   & \ptn   & \ptn   \\
 1.332 & 6.56 & 1.62 & 0.0734 & 0.0061 & 0.0385 \\
 1.457 & 7.45 & 1.47 & \ptn   & \ptn   & \ptn   \\
 1.586 & 7.62 & 1.10 & 0.0801 & 0.0062 & 0.0458 \\
 1.721 & 6.02 & 0.54 & \ptn   & \ptn   & \ptn   \\
 1.848 & 5.30 & 0.37 & 0.1108 & 0.0099 & 0.0742 \\
 1.995 & 4.36 & 0.20 & \ptn   & \ptn   & \ptn   \\
 2.101 & 3.41 & 0.11 & 0.1391 & 0.0115 & 0.0767 \\
 2.177 & 3.31 & 0.14 & \ptn   & \ptn   & \ptn   \\
 2.262 & 2.84 & 0.12 & \ptn   & \ptn   & \ptn   \\
 2.355 & 2.36 & 0.09 & 0.1666 & 0.0143 & \ptn   \\
 2.443 & 2.01 & 0.05 & \ptn   & \ptn   & \ptn   \\
 2.532 & 1.66 & 0.01 & \ptn   & \ptn   & \ptn   \\   \hline
\end{tabular}
\end{table}

\end{document}